\documentclass[aps,prd,floatfix,superscriptaddress,twocolumn,showpacs,amssymb]{revtex4-1}
\usepackage{amssymb,amsmath,verbatim}
\usepackage{graphicx}
\usepackage{epsfig}
\usepackage{dcolumn}
\usepackage{bm}

\newcommand{\be}{\begin{equation}}
\newcommand{\ee}{\end{equation}\noindent}
\newcommand{\eei}{\end{equation}}
\newcommand{\bea}{\begin{eqnarray}}
\newcommand{\eea}{\end{eqnarray}\noindent}
\newcommand{\eeai}{\end{eqnarray}}

\newcommand{\hf} {\frac{1}{2}}
\newcommand{\nn}{\nonumber\\}

\def\eq#1{(\ref{#1})}

\def\ord#1{{\cal O}(#1)}

\def\h#1{{\hat#1}}

\def\c#1{{\cal#1}}
\def\b#1{{\bar#1}}
\def\pa{\parallel}
\def\pe{\perp}

\def\Tr{{\mathrm{Tr}}}

\def\fdd#1#2#3{\frac{\delta^2#1}{\delta#2\delta#3}}
\def\fddd#1#2#3#4{\frac{\delta^3#1}{\delta#2\delta#3\delta#4}}
\def\fdddd#1#2#3#4#5{\frac{\delta^4#1}{\delta#2\delta#3\delta#4\delta#5}}

\begin{document}
\title{
Effect of the quartic gradient  terms on the critical exponents of the Wilson-Fisher fixed point in $O(N)$ models
}

\author{Z. P\'eli}
\affiliation{Department of Theoretical Physics, University of Debrecen,
P.O. Box 5, H-4010 Debrecen, Hungary}
\author{S. Nagy}
\affiliation{Department of Theoretical Physics, University of Debrecen,
P.O. Box 5, H-4010 Debrecen, Hungary}
\author{K. Sailer}
\affiliation{Department of Theoretical Physics, University of Debrecen,
P.O. Box 5, H-4010 Debrecen, Hungary}

\date{\today}

\begin{abstract}
The effect of the $\ord{\partial^4}$ terms of the gradient expansion on anomalous dimension $\eta$ and the correlation length's critical exponent $\nu$ of the Wilson-Fisher fixed point has been determined  for the  Euclidean $O(N)$ model for $N=1$ and the number of dimensions $2< d<4$ as well as
for $N\ge 2$ and   $d=3$.  Wetterich's effective average action  renormalization group method is used  with field-independent derivative couplings  and Litim's optimized regulator. It is shown that the critical theory for $N\ge 2$ is well approximated by  the effective average action preserving $O(N)$ symmetry
with the accuracy of $\ord{\eta}$.

\end{abstract}



\maketitle
\section{Introduction}

In the present paper it is investigated the effect of the term of $\ord{\partial^4}$ of the gradient expansion (GE) on the anomalous dimension $\eta$ and the correlation length's critical exponent $\nu$ at the Wilson-Fisher (WF) fixed point (FP)  for  Euclidean $O(N)$ models for $N=1$ and the number of dimensions $2< d<4$ as well as for $N\ge 2$ and   $d=3$. Wetterich's effective average action (EAA) renormalization group (RG) approach \cite{WePLB1993,Wette2002} is applied in the GE using field-independent (uniform) derivative couplings and Litim's optimized regulator \cite{Litim2001b}. The $N=1$ case for the number of dimensions $2<d<4$ is investigated in detail in order to show that the RG scheme used by us provides qualitatively reasonable results for the  exponents $\eta$ and $\nu$, even if we do not pretend to achieve an accuracy available only for taking the field-dependence of all couplings properly into account \cite{Wette2002,Canet2002,Canet2003}. Then we apply the same RG scheme  to $O(N)$ models with $N\ge 2$ and $d=3$ in order to show that the critical theory corresponding to the WF FP is well approximated by an EAA in which the derivative couplings for the radial  and the Goldstone modes of the field take different values and show up different $N$-dependences.  Such an EAA breaks $O(N)$ symmetry explicitly, but not more than  the order of magnitude of the anomalous dimension.  According to our knowledge the effect of the terms of $\ord{\partial^4}$   on the  exponents $\eta$ and $\nu$ has only be investigated for the 3-dimensional $O(1)$ model \cite{Canet2003} but not for $O(N)$ models with  $N\ge 2$ and $d=3$. Therefore even our qualitative statements on the $N$-dependence of the modification of the values of $\eta$ and $\nu$ due to the inclusion of the running higher-derivative term of $\ord{\partial^4}$ may fill this lack of informations.

 As a rule, the RG flow equations  can not be solved exactly, but their solution requires some  truncated approximation scheme. For the GE the leading order is the local potential approximation (LPA) when only the local potential evolves, whereas in the next-to-leading order (NLO) and the next-to-next-to-leading order (NNLO) also the couplings of the gradient terms of the quadratic and quartic orders are evolved,  respectively. In more advanced applications one takes into account the field-dependence of the derivative couplings (like e.g., in Refs.  \cite{Wette2002,Canet2002,Canet2003}), let us call below these approximation schemes NLOf and NNLOf, respectively. Here we restrict ourselves to the more simple NLOu and NNLOu approximation schemes, when the field-dependence of the derivative couplings is neglected. This means that in our RG scheme only the momentum-dependence of the wavefunction renormalization is taken into account truncated at the terms quartic in the gradient. Although we are aware of the loss of obtaining very accurate values of
 the critical exponents in that manner, the more simple form of the RG evolution equations in the NNLOu scheme is advantageous for making qualitative conclusions on the NNLO effect, i.e.,  on the effect of the inclusion of the terms of $\ord{\partial^4}$   on the critical exponents of the WF FP.  
 In Appendix \ref{app:over} we give a short overview of the more recent efforts on the determination of the anomalous dimension and the correlation length's critical exponent by means of various functional RG schemes. The summary of various results in Tables \ref{tab:critexplit} and \ref{tab:critexplitN} shows  
 that the NLOu values of the anomalous dimension are in the range of the various NLOf values, so that the usage of the uniform wavefunction renormalization
seems to be quite acceptable when one goes further and asks for the qualitative 
behaviour of the NNLO effect on the critical exponents. 

In many cases the critical exponents were evaluated by the functional RG with the neglection of the ${\dot R}$f effect, i.e., that
 of the terms containing the scale derivatives of the derivative couplings in the scale-derivative ${\dot R}_k$  of the cutoff function. Such an approximation is believed to be justified by the smallness of the anomalous dimension. 
 For the 3-dimensional $O(1)$ model we  separate the NNLO effect and the 
${\dot R}$f effect and show that they cause comparable modifications of the NLO results on the critical exponents, but it turns out that the  ${\dot R}$f effect weakens by an order of magnitude when the NNLO approximation is used, at least in the case of using uniform wavefunction renormalization and Litim's optimized regulator.
Therefore,  it is  justified to neglect the ${\dot R}$f effect when we investigate the $N$-dependence of the NNLO effect on $\eta$ and $\nu$ in the $O(N)$ models with $N\ge 2$. Otherwise the treatment of the RG evolution equations would become extremely
involved in the NNLO of the GE even if one restricts oneself to the usage of uniform wavefunction renormalization and generates the RG evolution equations for the various couplings  by means of a computer algebraic program.

Below we shall apply the EAA RG approach  by making an ansatz for the EAA in the NNLO of the GE. Applying the usual techniques of the GE, one has to split the $N$-component ($N\ge 2)$ field variable into a homogeneous background piece plus the quantum fluctuations. The  homogeneous background defines an arbitrary but fixed direction in the internal space and the fluctuating field can be split into radial and transverse modes, where the latter play the role of the Goldstone modes in the symmetry broken phase of the $O(N)$ model for $d=3$.
Then the projection of the evolution equation onto the subspaces of the radial and the Goldstone modes yields different evolution equations for their momentum-dependent wavefunction renormalizations. Therefore one is enforced to introduce different  momentum-dependent wavefunction renormalizations for the radial and the Goldstone modes. When the field-dependence of the wavefunction renormalizations is taken into account, the  gradient terms for the radial and the Goldstone modes can be distinguished by different field-dependences and the $O(N)$ symmetry of the EAA can be kept (like e.g. in Refs. \cite{Tetra1993,Wette2002}). Here we neglect the field-dependence of the wavefunction renormalization completely. Then we have to distinguish the radial and the Goldstone modes by means of their different scale-dependent derivative couplings. In this way the symmetry broken phase of the model shall be described by an EAA in which the gradient terms break $O(N)$ symmetry explicitly, while the potential keeps the symmetry. Such an EAA imitates spontaneous breaking of symmetry  partially by explicit symmetry breaking, although the nontrivial minimum of the potential remains an evolving parameter vanishing in the limit when the gliding scale $k$ goes to zero.  We shall show that when this explicit breaking of symmetry is minimized, it does not exceed the order of magnitude of the anomalous dimension for the radial mode.  Then ourasymetric ansatz for the EAA provides a reasonable framework to describe the scaling behaviour near the WF FP.

Our strategy consists of the following steps: the determination of the quantities characterizing the WF crossover region by solving the RG evolution equations for nearly critical trajectories, 
 the determination of the quantities characterizing the WF FP by solving the fixed-point equations, and finally  establishing that the corresponding quantities obtained in the first two steps agree with high accuracy. As mentioned above in the case of $O(N)$ models with $N\ge 2$ our ansatz for the gradient piece of the EAA breaks $O(N)$ symmetry explicitly. Therefore it enables one to introduce the ratio  $\h{z}$ of the wavefunction renormalization of the radial mode to that of the Goldstone modes, and the declination of the value of $\h{z}$ from $1$ measures the strength of that symmetry breaking. The above described strategy of 
the investigation of the WF FP is then  applied under the constraint $\h{z}(k)\equiv 1$, i.e., that the $O(N)$ symmetry of  the EAA is enforced on the NLO level. Then the agreement between the corresponding quantities characterizing the WF crossover region and the WF FP are established, although an inconsistency of the RG scheme is observed. Namely, the anomalous dimensions for the radial and the Goldstone modes of the field turn out to be different, so that the beta-function of the ratio $\h{z}$ does {\em not} vanish, but takes a value of the order of the anomalous dimension for the radial mode. Therefore, we repeat the whole procedure with evolving the ratio $\h{z}$. In the latter case the difference of the anomalous dimensions of the radial and the Goldstone modes determines the evolution of $\h{z}$ to critical values $\h{z}^*$ generally much larger than $1$ and the quantities characterizing the WF crossover region show up power-law dependences on the gliding scale $k$. Then 
the solution of the fixed-point equations for fixed $\h{z}^*$ are in agreement with the corresponding characteristics of the WF crossover region. Nevertheless,
we shall show that the above described inconsistency of the RG scheme may even be enhanced. However, the power-law behaviour of the various quantities characterizing the WF crossover region enables one to look for a version of the applied RG scheme when the inconsistency is minimized. The latter turns out be the RG scheme with fixed $\h{z}(k)=1$ and can be interpreted as the physically realistic one.

The paper is organized as follows. In Sect. \ref{eaarg} our ansatz for the EAA for the $O(N)$ model is given and the general form of the RG evolution equations for the local potential and the momentum-dependent wavefunction renormalizations are derived. This RG framework is applied to the $O(1)$ model in Sect. \ref{o1mod} where our  strategy of the determination of the quantities characterizing the WF FP is described. Particular emphasis is given to the determination of the NNLO and  ${\dot R}f$ effects for the 3-dimensional $O(1)$ model, and the dependence of the NNLO effect on the continuous dimension $d$ in the interval $2<d<4$ is discussed. The same RG framework and strategy are used to study the $N$-dependence of the NNLO effect for $N\ge 2$ in Sect. \ref{onmods}. It is pointed out that the EAA of the critical theory  preserves $O(N)$ symmetry with an accuracy of the order  of the anomalous dimension of the radial mode of the field. The results are summarized in Sect. \ref{sum}. Appendix \ref{app:over} contains a short overview of the most recent results on the determination of the anomalous dimension and the correlation length's critical exponent in the framework of the functional RG approach. Appendix  \ref{app:nnlo2e} contains the
RG evolution equations in the approximation NNLO$2\eta$ which takes the ${\dot R}$f effect into account and involves a quartic potential.

\section{Effective average action renormalization group approach}\label{eaarg}

Applying the EAA RG approach to the  $N$-component scalar field $\phi_x^a$  $(a=1,2,\ldots, N)$   one splits the EAA ${\b{\Gamma}}_k[\phi]=\Gamma_k[\phi] +\Delta \Gamma_k[\phi]$ into the reduced EAA (rEAA) $\Gamma_k[\phi]$ and the regulator piece
\bea
\Delta \Gamma_k[\phi]&=& \frac{1}{2}\int_x \phi_x^a \c{R}_{k,x,y}^{a,b} \phi_y^b,
\eea
where 
\bea
   \c{R}_{k,x,y}^{a,b} &=&  R_k^{a,b}(-\Box)\delta(x-y)
\eea
with the choice of a field-independent infrared  (IR) cutoff matrix $R_k^{a,b}(u)$. Here and below the formulas are casted into the form that the differential operators act always on the index $x$ and $k$ denotes the running cutoff.
The Wetterich equation (WE) for the rEAA ${\Gamma}_k$ is given as
\bea\label{we} 
\dot{\Gamma}_k &=& \hf \Tr \biggl( \lbrack\Gamma^{(2)}_k + \c{R}_k  \rbrack^{-1}{\dot {\c{R}}}_k \biggr),
\eea  
where  the dot over the quantities indicates the scale-derivative $k\partial_k$, $\Gamma^{(2)}_k $ is a shorthand for the second functional derivative matrix
$\Gamma_{k,x,y}^{(2)a,b}= \fdd{\Gamma_k[\phi]}{\phi_x^a }{\phi_y^b}$   $(a,b=1,2,\ldots, N)$. The trace is taken over a complete set of field configurations.
Now we  make an ansatz for the rEAA in the NNLO of the GE. Application of the usual GE techniques involves the split of the field $\phi_x^a=\Phi^a + \eta^a_x$ into  the homogeneous background piece $\Phi^a=\Phi e^a$ and the inhomogeneous fluctuating field $\eta_x^a$, where $e^a$ is an arbitrarily fixed unit vector in the internal space $(e^a e^a=1)$.  For later convenience we introduce the projectors $\c{P}_\parallel^{ab}=e^ae^b$ and $\c{P}^{ab}_\perp=\delta^{ab}-e^a e^b$ acting on the $N$-vectors of the internal space and define the field components $\phi_{\parallel x}^a=\c{P}^{ab}_\parallel \phi^b_x$ and $\phi_{\perp x}^a=\c{P}^{ab}_\perp \phi^b_x$ of the radial and the transverse modes, respectively. It is well-known that the transverse modes  play the role of the Goldstone modes in the symmetry broken phase of the model. The latter are absent for the case with $N=1$.
For the rEAA we make the NNLO ansatz
\bea\label{eaaon}
  \Gamma_k[\phi] &=& \hf \int_{x,y} \phi_{\parallel x}^a D^{-1}_{\parallel x, y}(-\Box) \phi_{\parallel y}^a\nn
&& + \hf \int_{x,y} \phi_{\perp x}^a D^{-1}_{\perp x, y}(-\Box) \phi_{\perp y}^a + \int_x U_k(\rho_x) 
\eea
with
\bea\label{invpro}
  D^{-1}_{A x, y}(-\Box) &=& \c{Z}_{A~k}(-\Box) \delta_{x,y}
= (- Z_{A k}\Box+ Y_{A k} \Box^2)\delta_{x,y}\nn
\eea
where $ \c{Z}_{A~k}(-\Box)$ are the momentum-dependent wavefunction renormalizations for the radial $(A=\parallel)$ and the Goldstone $(A= \perp)$ modes and $\rho_x=\hf \phi^a_x\phi^a_x$.
In the symmetry broken phase the $\rho$-dependence of the potential is parametrized as
\bea\label{lpsbr}
U_k(\rho_x) &=&u_0
+ \sum_{n=2}^{M}\frac{u_n}{n!}(\rho_x-\rho^*)^n
\eea
with the scale-dependent couplings $u_n$ ($n\ge 2$) and the position of the
 minimum of the potential at $\rho^*$,
while in the symmetric case, i.e., for $\rho^*=0$, the parametrization 
\bea\label{lpsym}
   U_k(\rho_x) &=&\sum_{n=0}^M \frac{v_n}{n!}\rho_x^n
\eea
with the scale-dependent coupling $v_n$ can be used. Therefore, the potential is approximated by a polynomial of degree $M$ of the $O(N)$-invariant variable $\rho_x$.   
  The parametrization \eq{lpsbr} can only be used in the symmetry broken phase, whereas the parametrization \eq{lpsym} can be used in both phases. For the symmetry broken phase and  the truncation $M=2$ the two parametrizations of the potential are related
 by $v_0=u_0+\hf u_2\rho^{*2}$, $v_1=-2u_2 \rho^*$, and $v_2=u_2$, but we shall use the parametrization \eq{lpsbr} in the symmetry broken phase. The mass squared of the elementary excitations at the minimum of the potential
 are then given as
$m^2_{SB}=2u_2 \rho^*$ and $m^2_{S}=v_1$ in the symmetry broken and symmetric phases, respectively.
The ansatz
\eq{eaaon} treats the radial and the transverse (Goldstone)  modes of the field 
separately. This explicit breaking of $O(N)$ symmetry  provides  the flexibility to our RG approach that in the symmetry broken phase the dynamics may govern the system to states in which the momentum-dependent wavefunction renormalizations for these modes evolve  differently with the gliding scale $k$, although one starts the evolution  with the initial condition
$\c{Z}_{\perp\Lambda}(-\Box)=\c{Z}_{\parallel\Lambda}(-\Box)$  at the ultraviolet (UV) scale $\Lambda$ ensuring unbroken $O(N)$ symmetry of the bare action. 
 In this manner spontaneous symmetry breaking may be  mimicked by an explicit one.

The ansatz \eq{eaaon} with Eqs. \eq{invpro}-\eq{lpsym} has been inserted into the WE \eq{we}, then  evolution equations derived for the couplings of the gradient terms and  those of the local potential  by using  usual GE techniques.
The radial and the Goldstone modes were split as $\phi_{\pa x}^a=\Phi e^a+\eta_{\pa x}^a$ and $\phi_{\pe x}^a=\eta_{\pe x}^a$, respectively. 
 Both sides of the WE have been functional Taylor-expanded in powers of the fluctuating fields $\eta_{\pa x}^a$, $\eta_{\pe x}^a$ and the evolution equations for the local potential $U_k (\rho)$ and the momentum-dependent wavefunction renormalizations $\c{Z}_{k\pa}(Q^2)$ and $\c{Z}_{k\pe}(Q^2)$ read off by
comparing  the zeroth order terms and the quadratic ones on both sides of the WE, respectively.  The explicit evaluation of the traces on the right-hand side of the WE have been performed in the momentum representation. Denoting by $Q_\mu$ the momentum of the Fourier modes of the fluctuating field, and by $p_\mu$ the loop-momentum appearing in the explicit expressions of the traces, the regulator matrix  has been specified as a block-diagonal one 
$R_k^{a,b}(p^2)=\sum_{A=\pa, \pe}R_{A k}(p^2)\c{P}^{a,b}_A$ choosing the regulator functions $R_{A k}(p^2)$  in the form of Litim's optimized regulator,
\bea 
R_{A k}(p^2) &=&\lbrack Z_{A k}(k^2-p^2)+ Y_{A k}(k^4-p^4)\rbrack\Theta(k^2 - p^2 )\nn
\eea
with the Heaviside function $\Theta(u)$. This choice reduces the loop-integrals appearing in the traces to those over the Euclidean sphere of radius $k$ and has the advantage that the loop-integrals can be taken analytically.

Truncating the  functional Taylor-expansion at the quadratic term,  the left-hand side of the WE  takes the form 
\bea\label{dotga} 
{\dot{\Gamma}}_k[\Phi+\eta] &=&{\dot{\Gamma}}_k[\Phi]
+\hf \int_{x,y} \eta_x^a {\dot A}_{k,x,y}^{a,b} \eta_y^b,
\eea
while the expansion of the right-hand side can be written as
\bea\label{ga2exp}
\Gamma_{k,x,y}^{(2)a,b}[\Phi+\eta] &=& A_{k,x,y}^{a,b} + (\eta B)_{k,x,y}^{a,b}+ \hf (\eta C\eta)_{k, x,y}^{a,b},
\eea
where
\bea\label{A}
  A_{k,x,y}^{a,b}&=& \Gamma_{k,x,y}^{(2)a,b}[\Phi],\\
\label{B}
 (\eta B)_{k,x,y}^{a,b}&=& \int_z  \eta_z^c \fddd{\Gamma_k}{\phi_x^a}{\phi_y^b}{\phi_z^c}\biggr|_{\phi_z=\Phi},\\
\label{C}
(\eta C\eta)_{k, x,y}^{a,b}&=&  \int_{z,u} \eta_z^c \fdddd{\Gamma_k}{\phi_x^a}{\phi_y^b}{\phi_z^c}{\phi_u^d}\biggr|_{\phi_z=\Phi}  \eta_u^d.
\eea
 The first-order term on the left-hand side vanishes because $\eta_x$ contains no zero mode. 
The field-independence of the gradient terms leads to the great simplification 
 that the third and fourth functional derivatives of the rEAA
come from the derivatives of the potential alone. The functional Taylor-expansion of the trace on the right-hand side of the WE is then achieved by
performing the truncated Neumann-expansion of the inverse matrix
$\lbrack\Gamma^{(2)}[\phi] + \c{R}_k  \rbrack^{-1}$ at the IR cutoff  propagator
\bea
\!\!\! \!\!\!G_{p,q}^{a,b} &=&([\Gamma^{(2)}[\phi_B] + \c{R}_k]^{-1} )^{a,b}_{p,q}= \sum_{A=\parallel,\perp} G_A(p^2) \c{P}_A^{a,b},
\eea
where
\bea
\!\!\! \!\!\!\!\!\!\! G_\parallel(p^2)&=& [\c{Z}_{\parallel~k}(p^2) +U'_k(r) +2rU''_k(r)+ R_{\pa ~k}(p^2)]^{-1} \!,
\eea
and
\bea
G_\perp(p^2) &=&  [\c{Z}_{\perp~k}(p^2) +U'_k(r)+ R_{\pe ~k}(p^2)]^{-1}
\eea
with $r=\hf \Phi^2$ 
are the propagators of the radial and the Goldstone modes, respectively. Here and in what follows the notation of the $r$-dependence of the propagators has been suppressed in order to make our formulas more transparent.
The trace on the right-hand side of the WE \eq{we} can then be rewritten as
\bea\label{neuexp}
 \Tr \biggl( \lbrack\Gamma^{(2)} + \c{R}_k  \rbrack^{-1}{\dot {\c{R}}}_k \biggr)
&=& T_0 + T_1 + T_{2B} + T_{2C},
\eea
 where
\bea\label{Ts}
 T_0 &=&  \Tr[G{\dot R}_k],\nn
 T_1 &=& -\Tr[G \cdot (\eta B)\cdot G {\dot R}_k],\nn
T_{2B} &=& \Tr[G \cdot (\eta B)\cdot G \cdot (\eta B)\cdot G {\dot R}_k],\nn
T_{2C} &=& -\hf\Tr[G \cdot (\eta C \eta)\cdot G {\dot R}_k].
\eea
(Here the dot `$\cdot$' indicates matrix product both in the external and the internal spaces.)
One finds $T_1=0$ because the background is homogeneous and $\eta_x$ exhibits no zero mode. The other terms are given as
\bea
 T_0 &=& V \sum_{A=\parallel,\perp} d_A \int_p  G_A(p^2){\dot R}_{A~k}(p^2),
\eea
\begin{widetext}
\bea
 T_{2B}    
&=& \int_{Q,p} 2r  \biggl\{ \biggl\lbrack
  [G_\parallel (p^2)]^2 G_\parallel(q^2)\delta^2(r){\dot R}_{\parallel k}(p^2)
+ (N-1)[G_\perp(p^2)]^2 G_\perp(q^2)\epsilon^2(r){\dot R}_{\perp k}(p^2)
 \biggr\rbrack_{q=Q-p}\eta_{\parallel Q}\eta_{\parallel -Q}\nn
&&
+\epsilon^2(r) \biggl\lbrack
 [G_\parallel (p^2)]^2G_\perp(q^2){\dot R}_{\parallel k}(p^2)
+ [G_\perp(p^2)]^2 G_\parallel(q^2){\dot R}_{\perp k}(p^2) 
 \biggr\rbrack_{q=Q-p}
\eta_{\perp Q}^a \eta_{\perp -Q}^a
\biggr\},
\eea
\bea
 T_{2C}
&=&-\hf \int_{Q,p} \biggl\{
 \biggl\lbrack  [G_\parallel (p^2)]^2\gamma(r){\dot R}_{\parallel k}(p^2)
+ (N-1)[G_\perp(p^2)]^2\delta(r){\dot R}_{\perp k}(p^2) 
\biggr\rbrack \eta_{\parallel Q}\eta_{\parallel -Q}\nn
&&
+ \biggl\lbrack
   [G_\parallel (p^2)]^2\delta(r){\dot R}_{\parallel k}(p^2)
+(N+1) [G_\perp(p^2)]^2\epsilon(r){\dot R}_{\perp k}(p^2) 
\biggr\rbrack\eta_{\perp Q}^a \eta_{\perp -Q}^a
\biggr\}\nn
\eea
\end{widetext}
with
the degeneracies $d_\parallel=1$ and $d_\perp=N-1$ of the radial and the transverse modes, respectively, and
\bea
 \gamma(r)&=& 4r^2U''''_k(r)+12rU'''_k(r)+3U''(r),\nn
 \delta(r)&=& 2rU_k'''(r)+U_k''(r),\nn
 \epsilon(r)&=&U_k''(r).
\eea
The comparison of the terms of the orders $\ord{\eta^0}$ and $\ord{\eta^2}$
on both sides of Eq. \eq{neuexp} results in the evolution equations 
\bea\label{eveqs1}
 {\dot \Gamma}_k [\phi_B] &=& \hf T_0,
\eea
and
\bea\label{eveqs2}
(\eta  {\dot A}_k \eta) &=& T_{2B}+T_{2C},
\eea
respectively, 
where both sides of Eq. \eq{eveqs2} are diagonal in the radial and the Goldstone modes. Therefore one finds  the  evolution equations 
\bea\label{evU}
  {\dot U}_k (r)&=& \hf \sum_{A=\parallel,\perp} d_A \int_p  G_A(p^2){\dot R}_{A~k}(p^2)
\eea
for the local potential,
\bea
\label{cZparev}
\lefteqn{
 {\dot {\c{Z}}}_{\parallel k} (Q^2)     }\nn
&=&\int_p 2r\biggl\{
\biggl\lbrack 
 [G_\parallel (p^2)]^2 G_\parallel(q^2)[ 2rU'''_k+3U''_k]^2{\dot R}_{\parallel k}(p^2) \nn
&&
+ (N-1) [G_\perp(p^2)]^2G_\perp(q^2)[U''_k]^2{\dot R}_{\perp k}(p^2) 
\biggr\rbrack_{q=Q-p} \nn
&&
-  [G_\parallel (p^2)]^3[ 2rU'''_k+3U''_k]^2{\dot R}_{\parallel k}(p^2)\nn
&&
-(N-1)  [G_\perp(p^2)]^3[U''_k]^2{\dot R}_{\perp k}(p^2) 
\biggr\}_{r=r^*}
\eea
for the momentum-dependent wavefunction renormalization of the radial mode, and
\bea
\label{cZperev}
 {\dot {\c{Z}}}_{\perp k} (Q^2)&=&\int_p \biggl\{
2r[U''_k]^2\biggl\lbrack
   [G_\parallel (p^2)]^2G_\perp(q^2){\dot R}_{\parallel k}(p^2)\nn
&&
+  [G_\perp(p^2)]^2 G_\parallel(q^2){\dot R}_{\perp k}(p^2) 
\biggr\rbrack_{q=Q-p}\nn
&&
- [G_\perp(p^2)]^2U''_k{\dot R}_{\perp k}(p^2)
\biggr\}_{r=r^*}
\eea
for  the momentum-dependent wavefunction renormalization of the Goldstone modes.
The last two equations have been obtained by making use of the first and second derivatives of Eq. \eq{evU} with respect to the  field variable $r$. The notation of the $r$-dependence has been suppressed in the equations for the momentum-dependent wavefunction
 renormalizations. Furthermore, the right-hand sides of Eqs. \eq{cZparev} and \eq{cZperev} should be taken at the minimum of the potential $r=r^*$ in accordance with the usage of  field-independent derivative couplings.
Since there are propagators  in Eqs. \eq{cZparev} and \eq{cZperev} taken at the momentum $p-Q$ where $p$ is the loop-momentum, one has to Taylor-expand both sides of these equations  in powers of $Q_\mu$ and make use of $O(d)$ symmetry when performing integrals of the types 
\bea\label{odreps}
\int_p p_\mu p_\nu f(p^2) &=& d^{-1}\delta_{\mu\nu} \int p^2 f(p^2) ,\nn
\int_p p_\mu p_\nu p_\kappa p_\lambda f(p^2) &=& \lbrack d(d+2)\rbrack^{-1}
 \int_p (p^2)^2 f(p^2).
\eea
  Then the comparison of the terms of the orders $\ord{Q^2}$ and $\ord{Q^4}$ on both sides of Eqs. \eq{cZparev} and \eq{cZperev} provide  the evolution equations for the various couplings $Z_{A~k}$ and $Y_{A~k}$, respectively. The introduction of the dimensionless quantities shall be discussed below separately for the cases $N=1$ and $N\ge 2$. The explicit forms of the evolution equations for the dimensionless couplings have been generated by computer algebra.

At this point one has to emphasize that the introduction of the homogeneous background field $\Phi^a$ pointing into an arbitrary, but fixed direction $e^a$ in the internal space leads necessarily to different diagonal derivative pieces  for
the radial $\eta_{\pa}^a$ and the Goldstone $\eta_\pe^a$ modes, and finally to evolution equations of different forms even if identical momentum-dependent wavefunction renormalizations $\c{Z}_\pa(Q^2)=\c{Z}_\pe(Q^2)$ (and identical cutoffs $R_{\pa k}(p^2)=R_{\pe k}(p^2)$) would have been assumed. In the latter case, however, Eqs. 
\eq{cZparev} and \eq{cZperev} would have been in contradiction. Therefore, one 
can not avoid the introduction of different momentum-dependent wavefunction renormalizations $\c{Z}_\pa(Q^2)\not=\c{Z}_\pe(Q^2)$ for the radial and the Goldstone modes. This breaks the $O(N)$ symmetry of the rEEA explicitly, but can be considered as a kind of bookkeeping the consequences of the existence of the nontrivial minimum of the potential at $r=r^*$ in the symmetry broken phase. Our ansatz
allowing for different RG evolutions of  $\c{Z}_\pa(Q^2)$ and $\c{Z}_\pe(Q^2)$
makes the RG scheme more flexible and raises the question that starting the evolution from a symmetric initial state with  $\c{Z}_\pa(Q^2)=\c{Z}_\pe(Q^2)$  at the UV scale, whether the critical theories  at the WF FP for $N\ge 2$  exhibit this symmetry or not.

The ansatz \eq{eaaon} with Eq. \eq{invpro} enables one to discuss various truncations of the GE: the LPA for $Z_{\pa k}=Z_{\pe k}\equiv 1$, $Y_{\pa k}=Y_{\pe k} \equiv 0$,  the NLO of the GE with scale-dependent wavefunction renormalizations  $Z_{\pa k}$, $Z_{\pe k}$ and $Y_{\pa k}=Y_{\pe k}\equiv 0$, whereas the running of all derivative  couplings $Z_{\pa k}$, $Z_{\pe k}$, $Y_{\pa k}$, and $Y_{\pe k}$ corresponds to the NNLO of the GE. These  truncations of the GE with given  $M$ of the truncation of the field-dependence of the local potential shall be referred to below as NLO$M$ and NNLO$M$ approximations, when the ${\dot R}$f effect is neglected. The notations NLO$M\eta$ and NNLO$M\eta$ refer to approximations when the  ${\dot R}$f effect was taken into account. Our results are obtained with uniform wavefunction renormalization, i.e., field-independent derivative couplings. This will not be indicated in the notation of the approximation except of the cases when 
 it should be emphasized with comparison of results from the literature obtained by the usage of either uniform (NLOu, NNLOu) or field-dependent (NLOf, NNLOf) derivative couplings. We have found that our results became numerically stable for the truncation $M=6$.

 \section{$O(1)$ model for dimensions $2<d<4$}\label{o1mod}

\subsection{Evolution equations}
Formally one has to set $N=1$, remove the evolution Eq. \eq{cZperev} for $\c{Z}_\pe(Q^2)$ and work with the derivative couplings $Z_k=Z_{\pa k}$ and $Y_k=Y_{\pa k}$. Then the evolution Eqs. \eq{evU} and \eq{cZparev} reduce to the equations
\bea\label{potflow}
   {\dot U}_k(r) &=& 
\hf \int_p G(p^2) {\dot R}_{ k }(p^2)
\eea
for the potential $U_k(r)$ and 
\bea\label{wfrflow}
\lefteqn{
 {\dot {\c{Z}}}_{ k}(Q^2) 
=
 2r^*[ 2r^* U_k'''(r^*) + 3 U''(r^*)]^2 }\nn
&&\times\int_p 
\biggl(
 G^2(p^2) [G(q^2)]_{q=Q-p }-  G^3(p^2)  \biggr)  {\dot R}_{ k}(p^2)
\eea
for  the momentum-dependent  wavefunction renormalization $\c{Z}_k(Q^2)$, where
$G(p^2)=G_\pa(p^2)$ and $R_k(p^2)=R_{\pa k}(p^2)$.

As to the next one introduces the dimensionless quantities
 $\b{r}= Z_k k^{-(d-2)} r$, $\b{r^*}=\b{\kappa}$, $\b{u}_n=
 Z_k^{-n}k^{-d+n(d-2)}u_n$, $\b{v}_n=  Z_k^{-n} k^{-d+n(d-2)}v_n$, and $\b{Y}_k =Z_k^{-1} k^2 Y_k$. These definitions of the dimensionless quantities incorporating  
appropriate powers of the uniform wavefunction renormalization is advantageous
 because they make disappear the coupling $Z_k$ from the beta-functions. 
 Then one
 arrives to the explicit form of the RG equations for the various couplings.
These have been generated by computer algebra for the various approximation schemes both for the symmetry broken and symmetric phases.
 
For the symmetry broken phase and the approximation scheme NNLO$2$  (with $\b{\lambda}=\b{u}_2$) the
evolution equations are given as
\bea\label{bkaNNLO}
\dot{\b{\kappa}}
&=&
-(d-2+\eta)\b{\kappa}+a(1+2\b{Y}_k)\b{g}^2\equiv\beta_{\b{\kappa}}
, 
 \eea
\bea\label{blaNNLO}
\dot{\b{\lambda}}
&=&
(d-4+2\eta)\b{\lambda}+b(1+2\b{Y}_k)\b{\lambda}^2\b{g}^3
\equiv\beta_{\b{\lambda}}
,
\eea
\begin{widetext}
\bea\label{bYNNLO}
\dot{\b{Y}}_k
&=&
(2 +\eta)\b{Y}_k+
18\alpha_d\b{\kappa}  \b{\lambda }^2  (1+ 2\b{Y}_k)\b{g}^4 
 \biggl\{
\biggl\lbrack
   \frac{48  }{d (d+2) (d+4)}
+\frac{576  \b{Y}_k }{d (d+2) (d+6)}
+\frac{192}{d(d+8)}\biggl( \frac{1 }{d}
+\frac{12  }{d+2 }\biggr)  \b{Y}^2_k \nn
&&
+ \frac{1280}{d(d+10)} \biggl( \frac{1  }{d}
+\frac{3 }{ d+2 }
\biggr) \b{Y}^3_k 
+\frac{1}{d(d+12)}\biggl( \frac{1792  }{d}
+\frac{6144 }{d+2} 
\biggr) \b{Y}^4_k 
\biggr\rbrack  \b{g}^3
\nn
&&
-
\biggl\lbrack
\frac{12}{d (d+2)} 
+\frac{40}{d(d+4)}\biggl(3+ \frac{2}{d}
\biggr)  \b{Y}_k 
+\frac{160}{d(d+6)}\biggl( 3+\frac{4}{d}
\biggr)  \b{Y}^2_k 
+\frac{192}{d(d+8)}\biggl( 5+\frac{6 }{d}
+\frac{12 }{d+2}  
\biggr) \b{Y}^3_k 
\biggr\rbrack  \b{g}^2
\nn
&&
+\biggl\lbrack
\frac{ 1}{d}
  +\frac{8}{d+2}\biggl(
1+\frac{6}{d  }
\biggr)\b{Y}_k
+\frac{24}{d+4}\biggl(1 +\frac{12  }{d}
\frac{2}{d^2 }
+\frac{6  }{d (d+2) }
\biggr) \b{Y}^2_k 
\biggr\rbrack\b{g} 
-\frac{3 \b{Y}_k }{d}
\biggr\} \equiv\beta_{\b{Y}}
,
\eea
\end{widetext}
where the anomalous dimension is given now as
\bea\label{betaNNLO}
 \eta&=& 
-36 \alpha_d\b{\kappa}  \b{\lambda }^2\biggl(1+ 2\b{Y}_k \biggr)
\b{g}^4
\biggl\{
\frac{4}{d}\biggl\lbrack 
\frac{ 1}{d+2}+
\frac{8  \b{Y}_k }{d+4}\nn
&&
+\frac{24   \b{Y}^2_k }{d+6}
\biggr\rbrack \b{g}
-\frac{1+6\b{Y}_k}{d}
\biggr\}\equiv \eta(\b{\kappa},\b{\lambda},\b{Y}_k),
\eea
and we have introduced
\bea\label{bgNNLO}
\b{g}&=& \lbrack 1 +\b{Y}_k+ 2 \b{\kappa}  \b{\lambda }\rbrack^{-1} 
\eea
and the constants $a= 6\alpha_d/d$, $b= 6a$, $\alpha_d=\hf\Omega_d (2\pi)^{-d}$ with the $d$-dimensional solid angle $\Omega_d$.
In order to determine the RG trajectories one has to solve first the coupled set of first order ordinary differential equations
\eq{bkaNNLO}-\eq{bYNNLO}
  for  initial conditions given at the UV scale $k=\Lambda$ 
by making use of Eq. \eq{betaNNLO}, and then determine the evolution of the wavefunction renormalization $Z_k$ by the integration of the equation
\bea\label{Zbeta}
  {\dot Z}_k&=& -\eta Z_k.
\eea
 The evolution equations for the more restrictive truncations of the GE can be obtained from the NNLO equations, as described above.

The truncation of the EAA with field-independent wavefunction renormalization $\c{Z}_k(Q^2)$ has the disadvantage that in the symmetric phase the wavefunction renormalization $Z_k$ does not evolve, keeps its UV value $Z_k=1$. The modification of the wavefunction renormalization in the symmetric phase occurs as a two-loop effect in the perturbative approach, the RG aproach reveals it only if the field-dependence
of the wavefunction renormalization is taken with. Therefore we shall concentrate our numerical work on the symmetry broken phase of the model. Then the scale-dependence of $Z_k$ may occur due to the nonvanishing value of $\kappa$, i.e., that of the nontrivial minimum of the potential.

 In order to calculate the most important  quantities characterizing the WF FP, we use the following strategy. First we solve the RG evolution equations for the  bunch $\c{B}$ of nearly critical trajectories running in the symmetry broken phase but in the close neighbourhood of the separatrix between the symmetric and symmetry broken phases of the model. The existence of the WF crossover scaling region shall be established in which these quantities take almost constant, i.e, scale-independent values,  $(\b{\kappa}_*, \b{\lambda}_*, \b{Y}_*)$, and
$\eta_*=\eta(\b{\kappa}_*, \b{\lambda}_*, \b{Y}_*)$. The closer the RG trajectory runs to the separatrix, the more pregnant is the crossover scaling region.
In order to control our procedure for the determination of the WF FP, we use the values found in the above described manner as educated guess for the solution  $(\b{\kappa}^*, \b{\lambda}^*, \b{Y}^*)$ of the fixed-point equations and finally establish that the solution of the latter is in good agreement with the `plateau' values. Then the anomalous dimension is redetermined as
$\eta^*=\eta(\b{\kappa}^*, \b{\lambda}^*, \b{Y}^*)$. As a byproduct, the lower end  of the WF scaling region  at the scale $k_c$ shall be used for the determination of the critical exponent $\nu$ in the manner described below.

\subsection{Crossover scaling}
\begin{figure}[htb]
\centerline{
\psfig{file=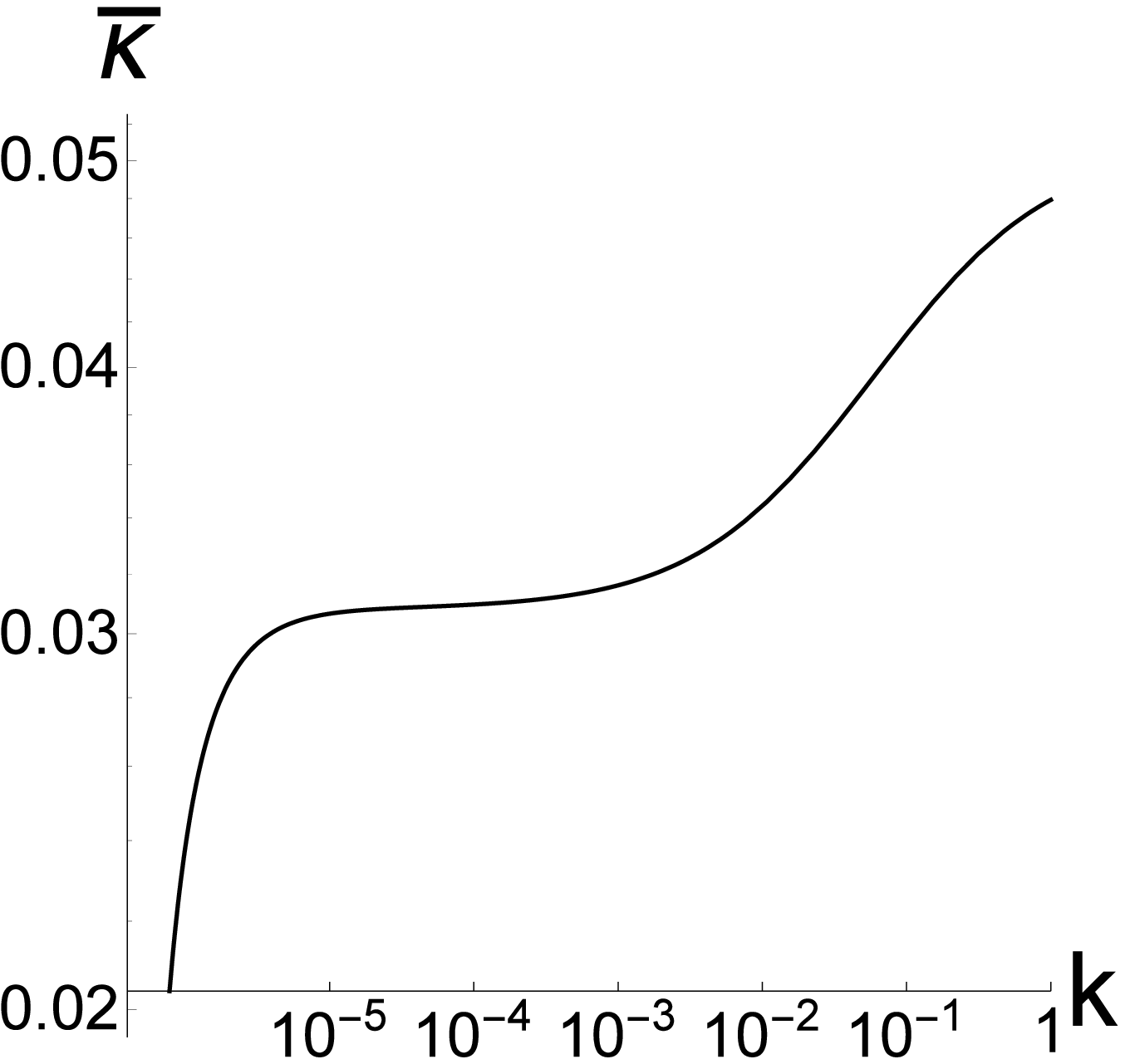,height=3.52cm,width=4.04cm,angle=0}
\psfig{file=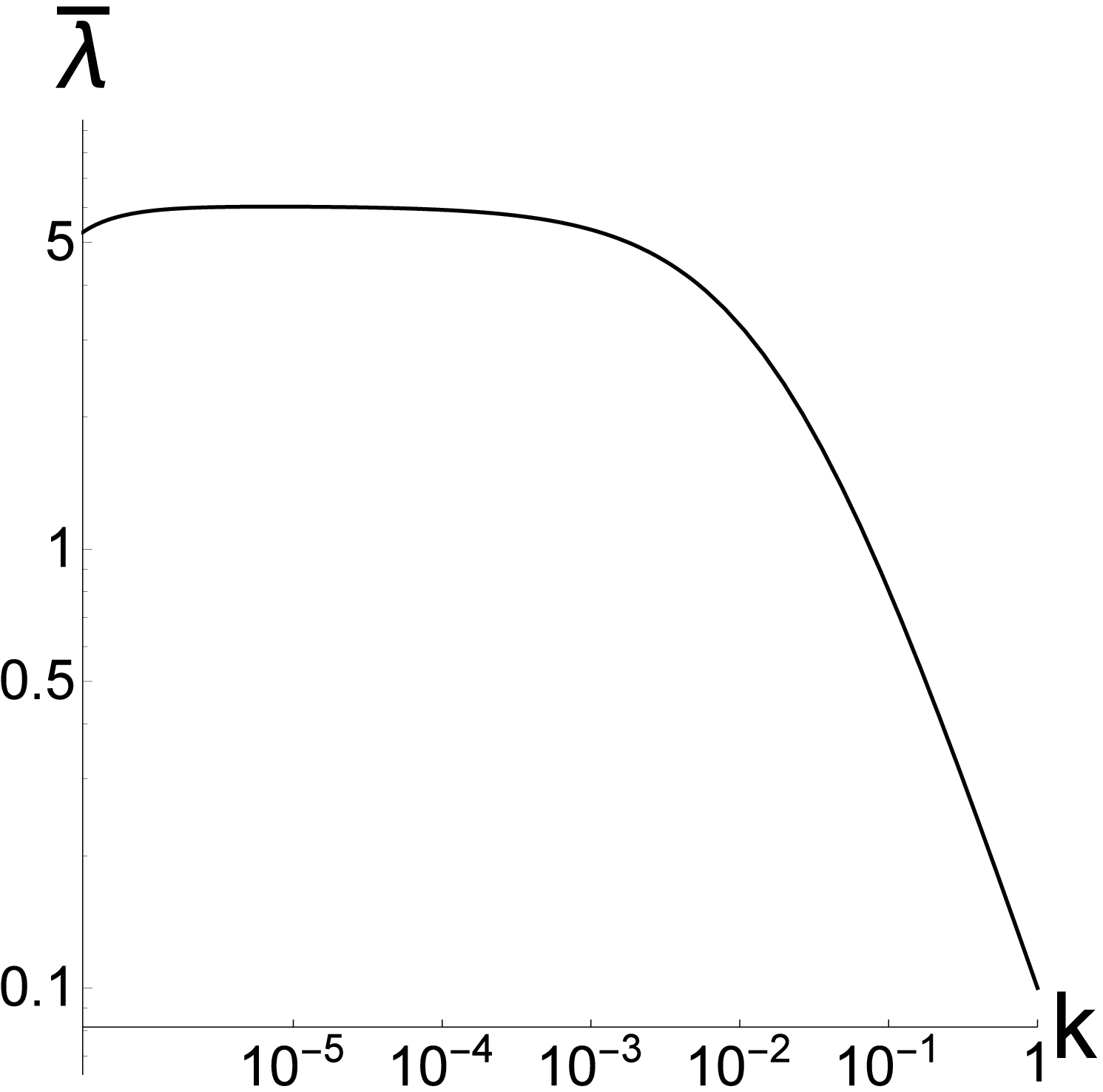,height=3.52cm,width=4.04cm,angle=0}
}
\centerline{
\psfig{file=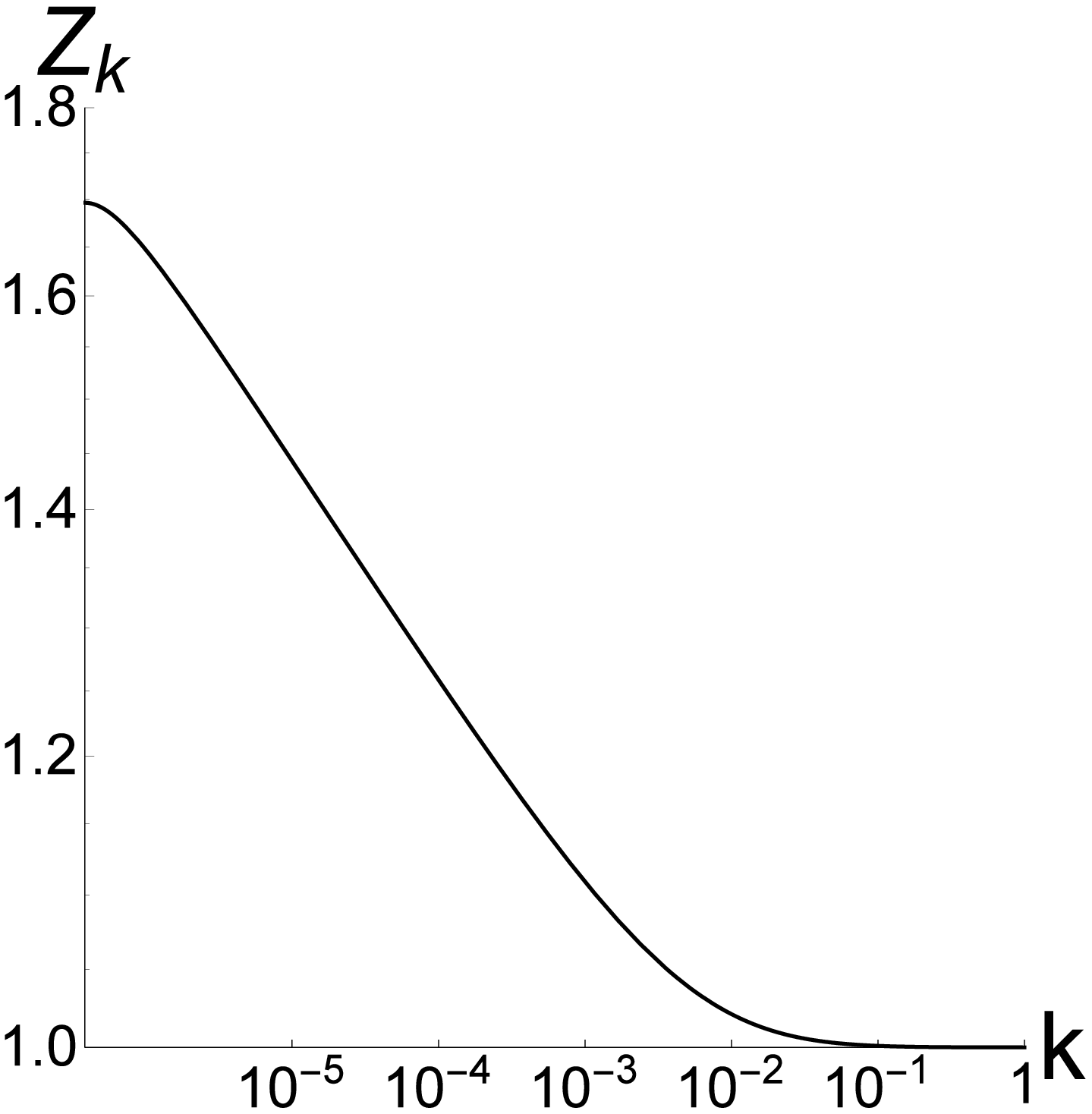,height=3.52cm,width=4.04cm,angle=0}
\psfig{file=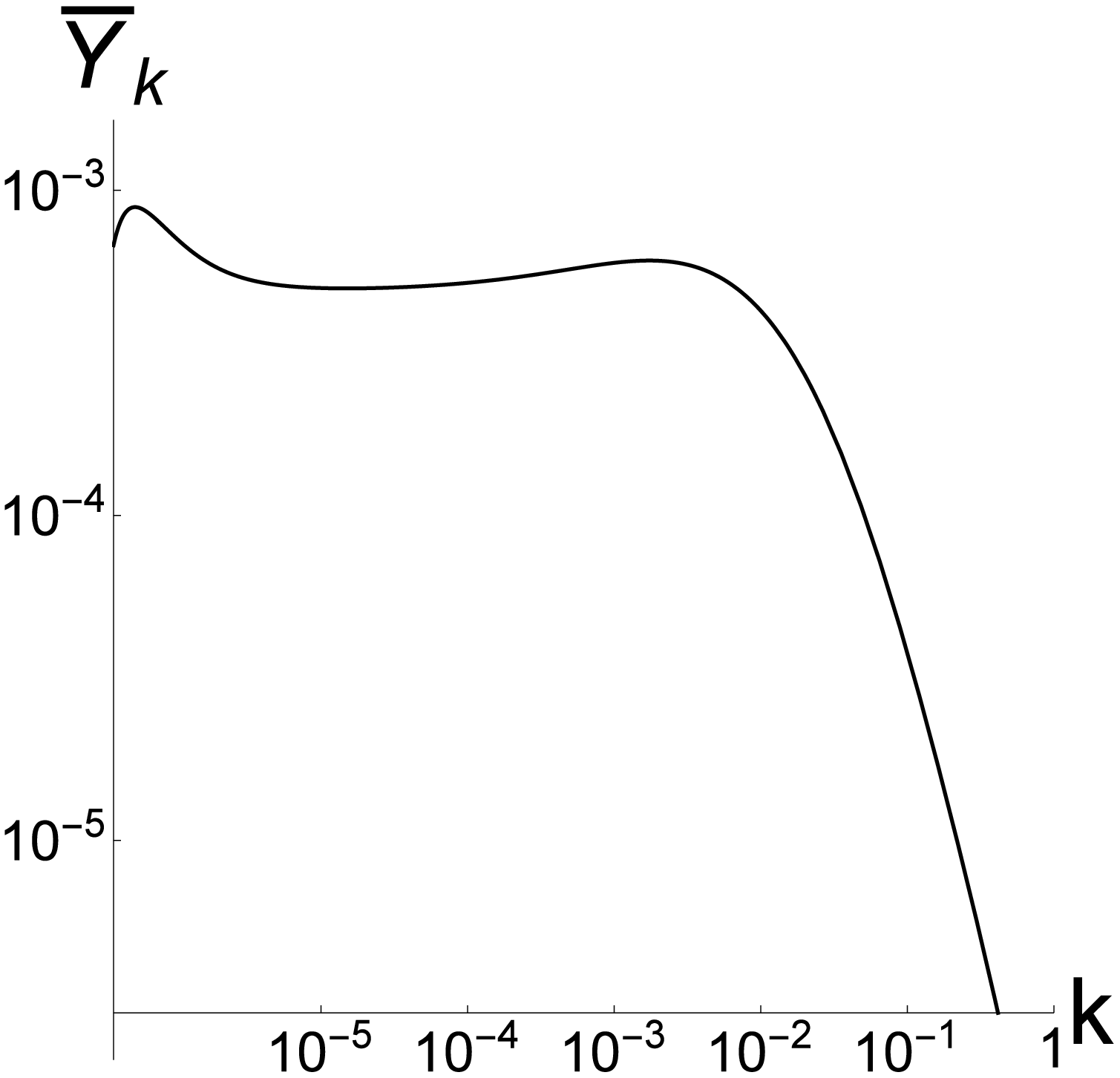,height=3.52cm,width=4.04cm,angle=0} }
\centerline{\psfig{file=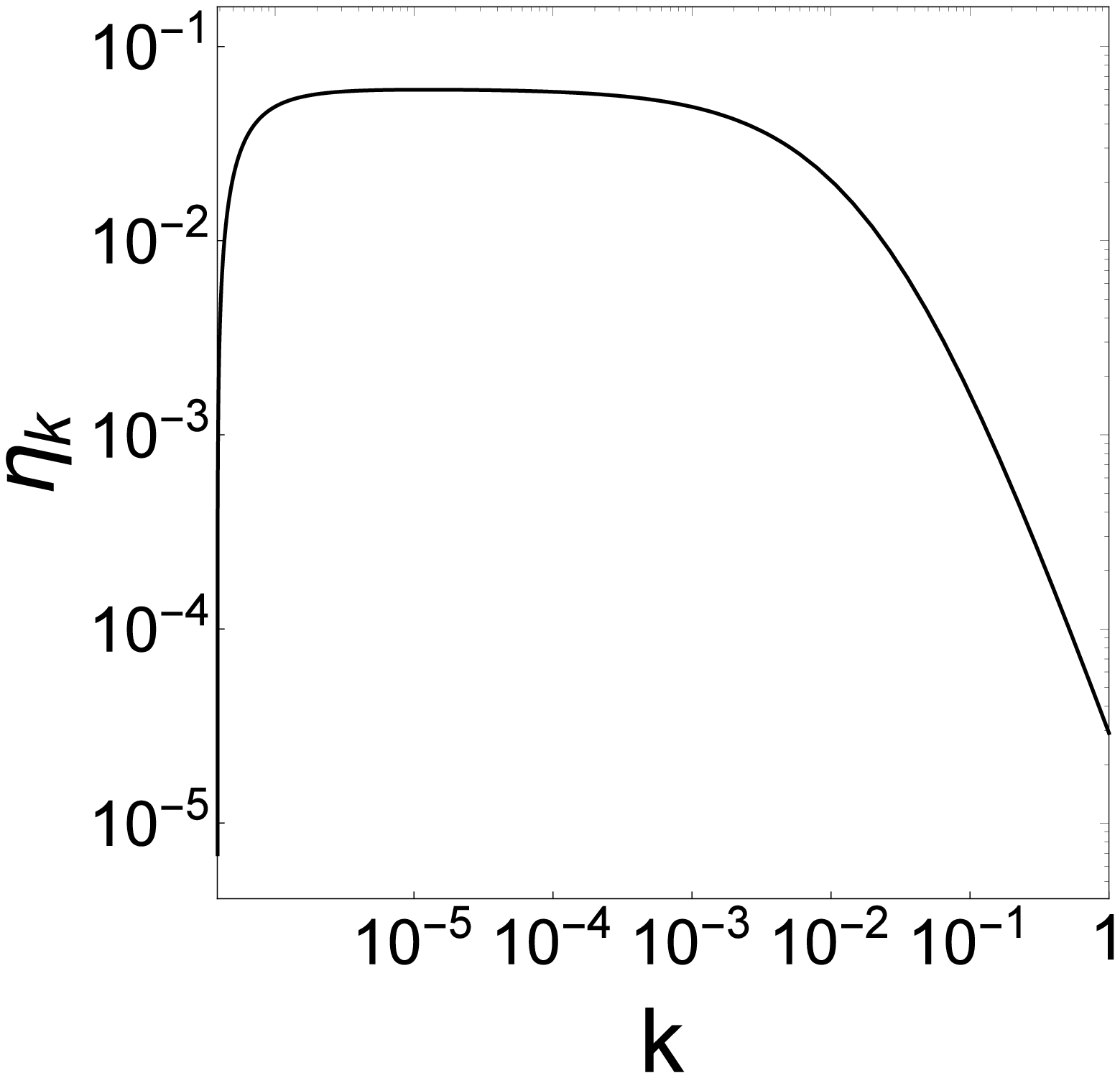,height=3.52cm,width=4.04cm,angle=0}
}
\caption{\label{fig:stdo2} Typical scale-dependence of  $\b{\kappa}$, $\b{\lambda}$, $Z_k$, $\b{Y}_k$, and $\eta$  on nearly critical RG trajectories with $Z_\Lambda=1$ and $\b{Y}_\Lambda=0$, determined in the approximation scheme  NNLO$6$ for $d=3$.}
\end{figure}

The quantities characterizing  the WF crossover region have been determined 
numerically by solving the RG evolution equations \eq{bkaNNLO}-\eq{bYNNLO} and \eq{Zbeta} (taking Eqs. \eq{betaNNLO} and \eq{bgNNLO} into account)  for a bunch $\c{B}$ of nearly critical RG trajectories.
The trajectories of bunch $\c{B}$ were chosen to belong to the UV couplings $\b{\lambda}_\Lambda=0.1$, $Z_\Lambda=1$, and $\b{Y}_\Lambda=0$.  The value $\b{\kappa}^{sep}_\Lambda$ identifying the separatrix was determined by fine tuning, by  looking for the RG trajectory on which the vanishing of $\b{\kappa}$ occurs at $k_c\sim 10^{-10}$, $\b{\kappa}(k_c)=0$.
The numerical investigation has been performed for various truncations $M$
 of the potential both in the NLO and NNLO of the GE, and it has been established that the truncation $M=6$ provides stable numerical values for the exponents $\eta$ and $\nu$. The typical evolution of the couplings $\b{\kappa}$, $\b{\lambda}$ ,$Z_k$, $\b{Y}_k$,  and the anomalous dimension $\eta$ are shown in Fig. \ref{fig:stdo2} for $d=3$. Similar behaviours were obtained for $d=4-\epsilon$ too.   Beyond the  short UV scaling region  
there occurs a rather long crossover scaling region stretched over  cca. 6 orders of magnitude change of the gliding scale $k$, in which $\b{\kappa}$ and $\b{\lambda}$ keep their constant values $\b{\kappa}_*$ and  $\b{\lambda}_*$, respectively. Also the function $\b{Y}_k$ turned out to be rather flat at its minimum with the value $\b{Y}_*$ and that flat region coincides with the position of the plateaus of the  functions  $\b{\kappa}(k)$ and $\b{\lambda}(k)$. 
Then even the anomalous dimension $\eta_*=\eta( \b{\kappa}_*,\b{\lambda}_*, \b{Y}_*)$ determined by means of Eq. \eq{betaNNLO} turns out to be scale-independent in the same region of the scale $k$. Since $\eta$ is constant in the scaling region,
the wavefunction renormalization $Z_k$ scales as $Z_k\sim k^{-\eta} $ in the WF crossover region.  All these  values characterizing the WF FP are  independent on the particular trajectory in the bunch $\c{B}$.

Beyond the WF crossover region the coupling $\b{\kappa}$  suddenly falls down to zero  at some finite scale $k_c$, $\b{\kappa}(k_c)=0$, signalling the lower end of the crossover region. Had been followed the evolution just on the separatrix, $k_c=0$ would be obtained. At the scale $k_c$ the  minimum of the potential is suddenly shifted to vanishing homogeneous background field, so that the symmetry of the vacuum state gets restored. The couplings $\b{\lambda}$ and $\b{Y}_k$  keep their finite nonvanishing values $\b{\lambda}_*$ and $\approx \b{Y}_*$, respectively, at the scale $k_c$, while
the wavefunction renormalization $Z_k$  goes to infinity on the separatrix and to some finite values on the various trajectories of the bunch $\c{B}$.
Below the scale $k_c$, i.e., in the  IR scaling region it holds $\kappa=0$,
 the anomalous dimension vanishes, and the field-independent wavefunction renormalization freezes out at its value $Z_{k_c}$ reached at the lower end $k_c$  of the WF crossover region, while the couplings of the potential and the coupling $\b{Y}_k$ show up tree-level scaling (see Eqs. \eq{bkaNNLO}, \eq{blaNNLO} and \eq{bYNNLO} for $\b{\kappa}=0$).

\begin{figure}[ht]
\centerline{\psfig{file=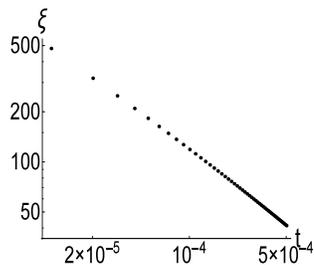,height=3.52cm,width=4.04cm,angle=0}}
\caption{\label{fig:nlonu2} Dependence of the  correlation length $\xi$ on the reduced temperature $t$ at the end of the WF crossover region in the approximation scheme NNLO6  for $d=3$.}
\end{figure}

The reciprocal of the scale $k_c$ can be identified with the correlation length 
$\xi =1/k_c$.  Let the separatrix be given by the initial conditions $(\b{\kappa}_\Lambda^{sep},\b{\lambda}_\Lambda, Z_\Lambda=1, \b{Y}_\Lambda=0)$. Then the 
  dependence of the correlation length $\xi$ on the
distance $\b{\kappa}_\Lambda^{sep}-\b{\kappa}_\Lambda=t^2$ of the particular RG trajectory with $(\b{\kappa}_\Lambda,\b{\lambda}_\Lambda, Z_\Lambda=1, \b{Y}_\Lambda=0)$ from the separatrix can be identified with the square of a kind of reduced temperature $t$ for any given initial value $\b{\lambda}_\Lambda$  \cite{Nagy2013}. It has been found that the correlation length scales with the reduced temperature as $\xi\propto t^{-\nu}$, where the critical exponent turned out to
be constant for the bunch of the trajectories $\c{B}$ (see Fig. \ref{fig:nlonu2} for that typical scaling behaviour). This qualitative behaviour is the same in the NLO and  NNLO of the GE for dimensions $d=3$ as well as $d=4-\epsilon$.

\subsection{Fixed points}

The fixed points are the solutions of
 the fixed-point equations
\bea\label{FPeqs}
 &&\beta_{\b{\kappa}}=\beta_{\b{\lambda}}=\beta_{\b{Y}}=0.
\eea
We shall pay particular emphasis on the determination of the WF FP, the solutions $(\b{\kappa}^*,\b{\lambda}^*,\b{Y}^*)$ of 
Eqs. \eq{FPeqs}.
 The wavefunction renormalization $Z_k$ does not occur in Eqs. \eq{FPeqs}, because it can be merged into the `renormalization' of the field variable for field-independent wavefunction renormalization. Therefore one should not require the vanishing of $\beta_Z$, instead of that Eq. \eq{betaNNLO} can be used to evaluate the value of the anomalous dimension 
 $\eta^*=\eta(\b{\kappa}^*,\b{\lambda}^*,\b{Y}^*) $
 at the FP. 

It is instructive to have a closer look on the fixed-point equations in the case  of the truncation $M=2$. Then the fixed-point equations $\beta_{\b{\kappa}}=\beta_{\b{\lambda}}=0$ can be solved analytically for any given $\b{Y}^*$, $\eta^*$, and dimension $d$,
 \bea\label{bkastar} 
\b{\kappa}^*(\b{Y}^*,\eta^*)
&=& \frac{2 \alpha_d}{3 d} \frac{(1+2\b{Y}^*) [4d + 5 (\eta^*-2)]^2}{(1+\b{Y}^*)^2 (d-2+\eta^*)^3},
\\
\label{blastar}
\b{\lambda}^*(\b{Y}^*,\eta^*)
&=&
-\frac{3 d}{2^2 \alpha_d} \frac{(1+\b{Y}^*)^3 (d-2+\eta^*)^3 (d-4 + 2 \eta^*)}{(1+2\b{Y}^*) [4d + 5 ( \eta^*-2)]^3}.\nn
\eea 
In the approximation  NLO$2$ one obtains the implicit equation
 $\eta^*=\eta(\b{\kappa}^*(\b{Y}^*=0,\eta^*),
\b{\lambda}^*(\b{Y}^*=0,\eta^*),\b{Y}^*=0)$ for the anomalous dimension which can be rewritten as
\bea\label{etastarNLO}
\lefteqn{
\eta_*
\stackrel{{\rm{NLO}}2}{=}
}\nn
&&
-\frac{(d+2\eta_* -4)^2 [ 3d^2 -16d +28+(16-3d)\eta_* ]}{
   9(d+2)(d-2+\eta_*)^2}.
\eea
Moreover,
the equations
\bea
0&=&\beta_{\b{Y}}(\b{\kappa}_*(\b{Y}_*,\eta_*), \b{\lambda}(\b{Y}_*,\eta_*),\b{Y}_*) 
\eea
and
\bea
  \eta_*&=&\eta( \b{\kappa}_*(\b{Y}_*,\eta_*), \b{\lambda}(\b{Y}_*,\eta_*),\b{Y}_*) 
\eea 
represent a system of equations for the determination of  $\b{Y}_*$ and $\eta_*$ in the approximation NNLO$2$. 

We restricted our discussion to dimensions $2<d<4$ and to the parameter region  $Z_k>0$, $\b{Y}_k\ge 0$, $\b{\lambda}\ge 0$ excluding unphysical cases such as the trivial fixed-point
 solution $Z_k=\b{Y}=\b{\lambda}= \b{\kappa}=0$ and those with Euclidean action
 unbounded from below. The fixed-point equations \eq{FPeqs} have a solution  with   $\b{\lambda}^*_G=0$ and $\b{Y}^*_G=0$ implying $\eta^*=0$  and $\b{\kappa}_G^*=a$ for any fixed value of $Z_k=Z$. Clearly, this represents the Gaussian FP in the LPA when the wavefunction renormalization is restricted to the value $Z_k=1$. In the approximation schemes NLO and NNLO there exists rather a Gaussian fixed line.

Eqs. \eq{FPeqs} were solved in various approximation schemes of the GE with a routine  which employs the Newton-Rhapson method. Its advantage is that it converges rapidly because the roots are calculated from gradients avoiding the calculation of numerical derivatives. The method needs rather good guesses as initial conditions for the roots. For these guesses we have taken the 'plateau' values $(\b{\kappa}_*,\b{\lambda}_*,\b{Y}_*)$ read off from the crossover scaling.

\subsection{Numerical results}

\subsubsection{Dimension $d=3$}

The above described numerical method yielded the
results for the position of the WF FP, the anomalous dimension $\eta^*$, and   the critical exponent 
$\nu$  as listed in Table \ref{tab:critexp} for various approximation schemes. 
\begin{table}[htb]
\begin{center}
\begin{tabular}{|c|c|c|c|c|c|}
\hline
 Approximation & $\b{\kappa}_*$ & ${\b{\lambda}}_*$ & $\b{Y}_*$ & 
$\eta_*$  & $\nu $\cr 
\hline\hline
  NLO2 & $0.025$ & $7.94$ & $-$ & $0.0537$ & $0.554$\cr
 NLO6 & $0.027$ & $6.23$ & $-$& $0.057$ & $0.615$\cr
 NLOu    \cite{Tetra1993}   &$0.041$ & $9.25$& $-$ & $0.045$    & $0.638$  \cr
 NLOf \cite{Canet2002}&$$  & $$& $-$&$0.044$& $0.628$ \cr
\hline
 NNLO2 & $0.025$ & $7.94$ & $0.0003$ & $0.0538$ & $0.560$  \cr
NNLO6 & $0.031$ & $6.02$ & $0.0005$ & $0.059$  & $0.634$  \cr 
NNLOf\cite{Canet2003}&  & &  &$0.033$ &$0.632$ \cr
\hline
\end{tabular}
\end{center}
\caption{\label{tab:critexp} The values of the various couplings, the anomalous dimension $\eta_*$, and the critical exponent $\nu$ characterizing the WF scaling region for $d=3$, obtained  in various approximation schemes. For comparison
the NLOu results taken from \cite{Tetra1993} and 
the NNLOf results taken from  Refs. \cite{Canet2002,Canet2003} are also shown.
}
\end{table}
 The values $\b{\kappa}_*$, $\b{\lambda}_*$, $\b{Y}_*$ (and correspondingly $\eta_*$) determined in this manner are consistent with the fixed-point Eqs. \eq{FPeqs}. Namely, inserting the  values $\eta_*$ and $\b{Y}_*=0$ for NLO2 and those of $(\eta_*, \b{Y}_*)$ for NNLO2 given in Table  \ref{tab:critexp} into Eqs.
\eq{bkastar} and \eq{blastar} reproduce the values $\b{\kappa}_*$ and $\b{\lambda}_*$ in Table  \ref{tab:critexp}  with high accuracy. Moreover,  Eq. \eq{etastarNLO} exhibits the numerical solution
$\eta^*=0.0537$ which agrees with the value found for NLO2 (see the first raw in Table  \ref{tab:critexp}).  Similar agreement has been found  for the truncation $M=6$ too. Therefore below we do not make any distinction between the dynamically obtained values $\b{\kappa}_*$, $\b{\lambda}_*$, etc. and the values  $\b{\kappa}^*$, $\b{\lambda}^*$, etc. obtained as the solution of the fixed-point equations.  

One recognizes from Table \ref{tab:critexp} that the inclusion of higher-order polynomial terms of the potential has rather significant effect on the position of the WF FP as well as on the values of the critical exponents, as expected. It is found that the WF FP is characterized by a nonvanishing value $\b{Y}^*$ of the higher-derivative coupling $\b{Y}_k$; this UV irrelevant coupling becomes relevant at the WF FP. 
 Our NLOu results for $\eta$ and $\nu$ obtained with Litim's optimized regulator overestimate those obtained in   \cite{Tetra1993} with the use of an exponential regulator, although the  ${\dot R}$f effect has been neglected in both cases, so that the difference has to occur due to the use of different regulators. 
 According to our NLO6u and NNLO6u  results the NNLO effect increases the values of $\eta$ and $\nu$ by $\sim$ 2 per cents. The comparison of the NLOu result from Ref.  \cite{Tetra1993} and the NLOf results from Ref. \cite{Canet2002} (obtained also with the use of the exponential regulator and with the full scale-derivative of the regulator function) shows together the ${\dot R}$f effect and the effect  that the inclusion of the  field-dependence of the wavefunction renormalization make out: these together alter the values of $\eta$ and $\nu$ by not more than $\sim$ 10 and $\sim 2$ per cents, respectively. When the field-dependence of the derivative couplings as well as the ${\dot R}$f effect are taken properly into account, the NNLO effect seems to  be of $\sim 30$ and $\sim 0.6$ per cents on $\eta$ and $\nu$, respectively, as the comparison of the NLOf\cite{Canet2002} and NNLOf\cite{Canet2003} data shows.
 Our estimates of the NNLO effect (obtained with uniform wavefunction renormalization and the neglection of the ${\dot R}$f effect) differ significantly from those observed in the NLOf\cite{Canet2002} and NNLOf\cite{Canet2003} data and our  NNLO6u values of $\eta$ and $\nu$ are overestimated as compared to
the NNLOf\cite{Canet2003} value. This may be the consequence of using Litim's regulator, neglecting the ${\dot R}$f effect and neglecting the field-dependence of the derivative couplings.

In order to separate the NNLO  and the  ${\dot R}$f effects when uniform wavefunction renormalization is used, we performed the determination of the critical exponents $\eta$ and $\nu$ in the framework of the NNLO$M\eta$ approximations too.
The full expression of the scale-derivative of the cutoff function can be rewritten as
\bea\label{rdoteta}
  {\dot R}_k&=&Z_k k^2 \biggl\lbrack 2 - \eta(1-y) + 2\b{Y}_k (1+y^2)\nn
&&
+ ({\dot{\b{Y}}}_k-\eta \b{Y}_k) (1-y^2) \biggr\rbrack, 
\eea
where $y=p^2/k^2$, while neglecting the scale-derivatives of the derivative couplings reduces this to 
\bea\label{rdot}
 {\dot R}_k &=&Z_k k^2 ( 2 + 4\b{Y}_k).
\eea
The expressions \eq{rdot} and \eq{rdoteta} have been used in the approximations
 NNLO$M$ and NNLO$M\eta$, respectively. The corresponding expressions for the approximations NLO$M$ and NLO$M\eta$ were obtained by setting $\b{Y}_k=0$ in expressions   
 \eq{rdot} and \eq{rdoteta}, respectively. The evolution equations for the dimensionless couplings become rather involved in the approximations NNLO$M\eta$. This is illustrated  in Appendix \ref{app:nnlo2e}, where the NNLO$2\eta$ analogues of evolution Eqs. \eq{bkaNNLO}-\eq{betaNNLO} are given. Our numerical results obtained in the approximations NLO$6\eta$ and NNLO$6\eta$ are given in Table \ref{tab:o1critexp}.
 One can see that (i) the ${\dot R}$f effect  at the NLO level  results in increments of $\sim 4$  per cents of the NLO values of $\eta$ and $\nu$ (compare the values in the third and first raws of Table \ref{tab:o1critexp}), which is comparable to the NNLO effect when the ${\dot R}$f effect is neglected (compare the values in the second and first raws of Table \ref{tab:o1critexp});   (ii) but the   ${\dot R}$f effect  at the NNLO level becomes much weaker and results in increments  of $\sim 0.3$ and $\sim 0.6$ per cents of the values of $\eta$ and  $\nu$, respectively (compare the values in the fourth and second raws of Table \ref{tab:o1critexp}). Thus we have found that the ${\dot R}$f effect becomes less significant in the NNLO approximation as compared to the NLO approximation, at least when uniform wavefunction renormalization and Litim's optimized regulator are used. This also means that the discrepancy between our NNLO$6\eta$ values and the NNLOf values in \cite{Canet2003} should to be prescribed to the regulator-dependence and the neglection of the field-dependence of the derivative couplings.

\begin{table}[htb]
\begin{center}
\begin{tabular}{|c|c|c|c|c|c|}
\hline
 Approximation & $\b{\kappa}^*$ & ${\b{\lambda}}^*$ & $\b{Y}^*$ & 
$\eta$  & $\nu $\cr 
\hline\hline
 NLO6       & $0.027$  & $6.23$ & $-$ & $0.0570$   & $0.6150$   \cr 
NNLO6       & $0.031$  & $6.02$ & $0.0005$  & $0.0590$ &$0.6340$ \cr
 NLO6$\eta$ & $0.031$ & $6.12$ & $-$ & $0.0590$  & $0.6377$ \cr 
 NNLO6$\eta$ & $0.031$ & $6.11$ & $0.0005$&  $0.0592$  & $0.6379$  \cr 
\hline
NNLOf\cite{Canet2003}&  & &  &$0.033$ &$0.632$ \cr
\hline
\end{tabular}
\end{center}
\caption{\label{tab:o1critexp} The ${\dot R}$f effect on the anomalous dimension $\eta$ and the correlation length's critical exponent $\nu$ for the 3-dimensional $O(1)$ model.
}
\end{table}

\subsubsection{Dimension $d=4-\epsilon$}

For dimension $d=4-\epsilon$ we restricted our discussion to the case of quartic potentials, i.e., to the truncation $M=2$. The characteristics of the WF FP have been determined in the first nonvanishing orders of the $\epsilon$-expansion analytically on the base of the fixed-point equations \eq{FPeqs} and numerically by identifying the WF crossover region along the RG trajectories of bunch $\c{B}$.  The fixed-point Eqs. \eq{FPeqs} can be used to find the first nonvanishing correction to the coupling of the $\ord{\partial^4}$ derivative term in the $\epsilon$-expansion.
 The analytical solution of  Eqs. \eq{FPeqs}  was determined with the accuracy
 of the  first nonvanishing $\epsilon$-dependent corrections for the truncation $M=2$ of the potential,
\bea\label{exps}
\!\!\!\! && \b{\kappa}^*=\kappa_0+ \kappa_1\epsilon +\ldots,~~~~~~~~
  \b{\lambda}^*=\lambda_0+\lambda_1 \epsilon+\ldots,\nn
\!\!\!\!&&  \b{Y}^*=\sum_{n=0}^3\frac{1}{n!}Y_n\epsilon^n+\ldots,~~~~
  \eta^*=\eta_0+\eta_1\epsilon +\hf \eta_2\epsilon^2.
\eea
Here the truncations of the $\epsilon$-expansions were dictated by the order-by-order successive solution of the fixed-point equations. 
 The vanishing of the zeroth-order terms of the beta-functions provides the zeroth-order solution $\lambda_0=0$ (implying $\eta_0=0$ since $\eta_0\propto \kappa_0\lambda_0^2$), $Y_0=0$, and
$\kappa_0=\frac{3}{2}\alpha_4 (1+2Y_0)(1+Y_0)^{-2}(2+\eta_0)^{-1}= \frac{3}{4}\alpha_4$. Inserting these results into the equations obtained by requiring the vanishing of the first-order terms of the beta-functions one finds nonvanishing values for $\kappa_1$ and $\lambda_1$ and
 again $Y_1=0$ and $\eta_1=0$ since those are proportional to $\kappa_0\lambda_1^2\epsilon^2$. Making use of the first-order results we find $\eta_2=1/18$ and surprisingly $Y_2=0$ because the expression in the curled bracket on the right-hand side of Eq. \eq{bYNNLO} vanishes in zeroth order of $\epsilon$, so that actually the first nonvanishing correction to the higher-derivative coupling
is of the order $\ord{\epsilon^3}$. Thus we find for the characteristics of the FP
\bea\label{4meps}
&&{\b{\kappa}}^*= \frac{3}{4} \alpha_4 \biggl\lbrack 1 
+\hf\biggl( \frac{11}{3} -2\gamma+ \ln (16\pi^2)\biggr)\epsilon
\biggr\rbrack,\nn
&&{\b{\lambda}}^*= \frac{16\pi^2}{9}\epsilon,~~
\b{Y}^*= \frac{1}{6\cdot 192}\epsilon^3,~~
\eta^*=\frac{1}{36}\epsilon^2
\eea
with the Euler-Mascheroni constant  $\gamma\approx 0.577$.  These analytic results are summarized in the second coloumn of Table
\ref{tab:fpquan4}.
\begin{table}[htb]
\begin{center}
\begin{tabular}{|c|c|c|c|}
\hline
 Quantity & Analytic result & Numerical result \cr
\hline\hline
$\b{\kappa}^*$ & $0.0047+ 0.009\epsilon$ &$ 0.0048+0.01\epsilon$\cr
$\b{\lambda}^*$& $17.5\epsilon$ & $ 16.5\epsilon$ \cr
$\b{Y}^*$& $ 8.7\times 10^{-4}\epsilon^3$ & $ 8.9\times 10^{-4}\epsilon^3$ \cr
$\eta^*$ & $0.028\epsilon^2$ &  $0.028\epsilon^2$ \cr
$\nu$   & $0.5+0.083\epsilon$\cite{Klein2001} & $0.51+0.095\epsilon$ \cr 
\hline
\end{tabular}
\end{center}
\caption{\label{tab:fpquan4} Position of the WF FP in the parameter space $(\b{\kappa},\b{\lambda},\b{Y})$ as well as the anomalous dimension $\eta^*$ and the critical exponent $\nu$ estimated  for quartic potentials (truncation $M=2$) in the NNLO of the GE for dimension $d=4-\epsilon$ (for $\epsilon\ll 1$). The two-loop analytical result for $\nu$ is taken from \cite{Klein2001}.}
\end{table}
The result for $\b{\lambda}^*$ agrees with the two-loop perturbative  result in 
\cite{Klein2001} (see Eq. \eq{2loop} and the note following it on the various definitions of the quartic coupling), while the two-loop result for the anomalous dimension differs of our result  by the factor $2/3$. Since our Eq. \eq{betaNNLO} for the anomalous dimension is the NNLO generalization of the NLO Eq. (5) in
\cite{Wette200310}, this discrepancy is independent of the RG scheme. It reflects that the EAA RG method sums up an infinite number of loop corrections in a nonperturbative manner. On the base of our analytic analysis we can also establish that the WF FP is characterized by a nonvanishing coupling
of the $\ord{\partial^4}$ term which turns out to be of the order $\epsilon^3$ in the $\epsilon$-expansion. This also means that the inclusion of the running coupling $\b{Y}_k$ into the gradient expansion does not affect the leading-order terms  of
$\b{\kappa}^*$, $\b{\lambda}^*$, and $\eta^*$ in their $\epsilon$-expansion.

The characteristic quantities of the WF crossover region were also determined numerically for several (a number of 100) values of $\epsilon$ taken with equal logarithmic distances in the interval $[10^{-7}, 10^{-1}]$.  The solution of the RG evolution equations as well as the identification of the scale $k_c$ were  performed  for the bunch $\c{B}$ of trajectories for each of these $\epsilon$ values. Then the $\epsilon$-dependence   of the various quantities was  fitted by polynomials of degrees given in Eq. \eq{exps} and by the linear relation $\nu=\nu_0+\nu_1\epsilon$. The characteristics of the WF FP obtained analytically and numerically are gathered  in Table \ref{tab:fpquan4}.
We see that the numerical results are in good agreement with the analytical
ones, and the same holds for
our numerical result and the two-loop result presented in \cite{Klein2001}
for the critical exponent $\nu$.

\subsubsection{Dependence on the continuous dimension $d$}

\begin{figure}[ht]
\centerline{\psfig{file=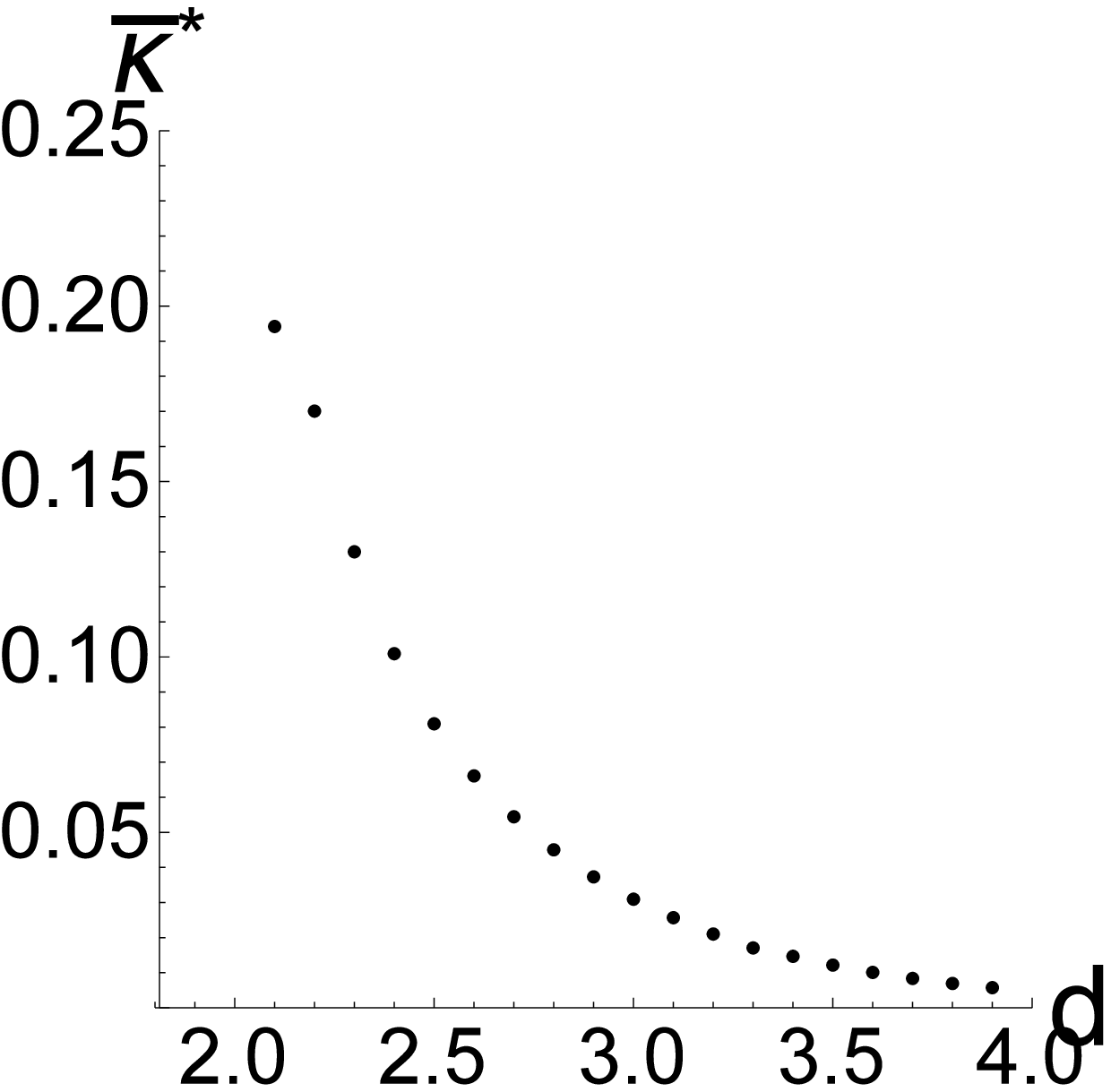,height=3.52cm,width=4.04cm,angle=0}
\psfig{file=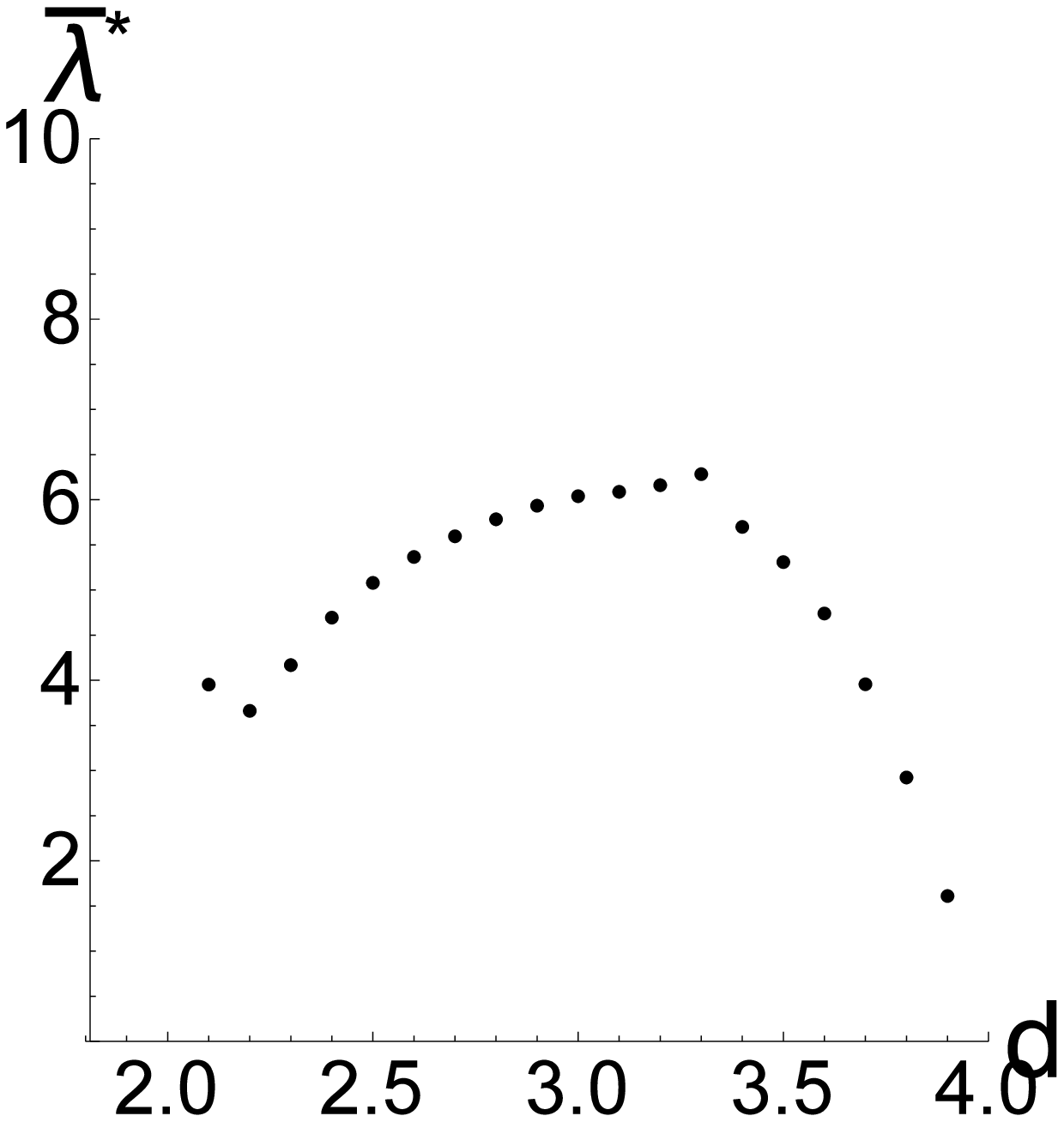,height=3.52cm,width=4.04cm,angle=0}}
\centerline{\psfig{file=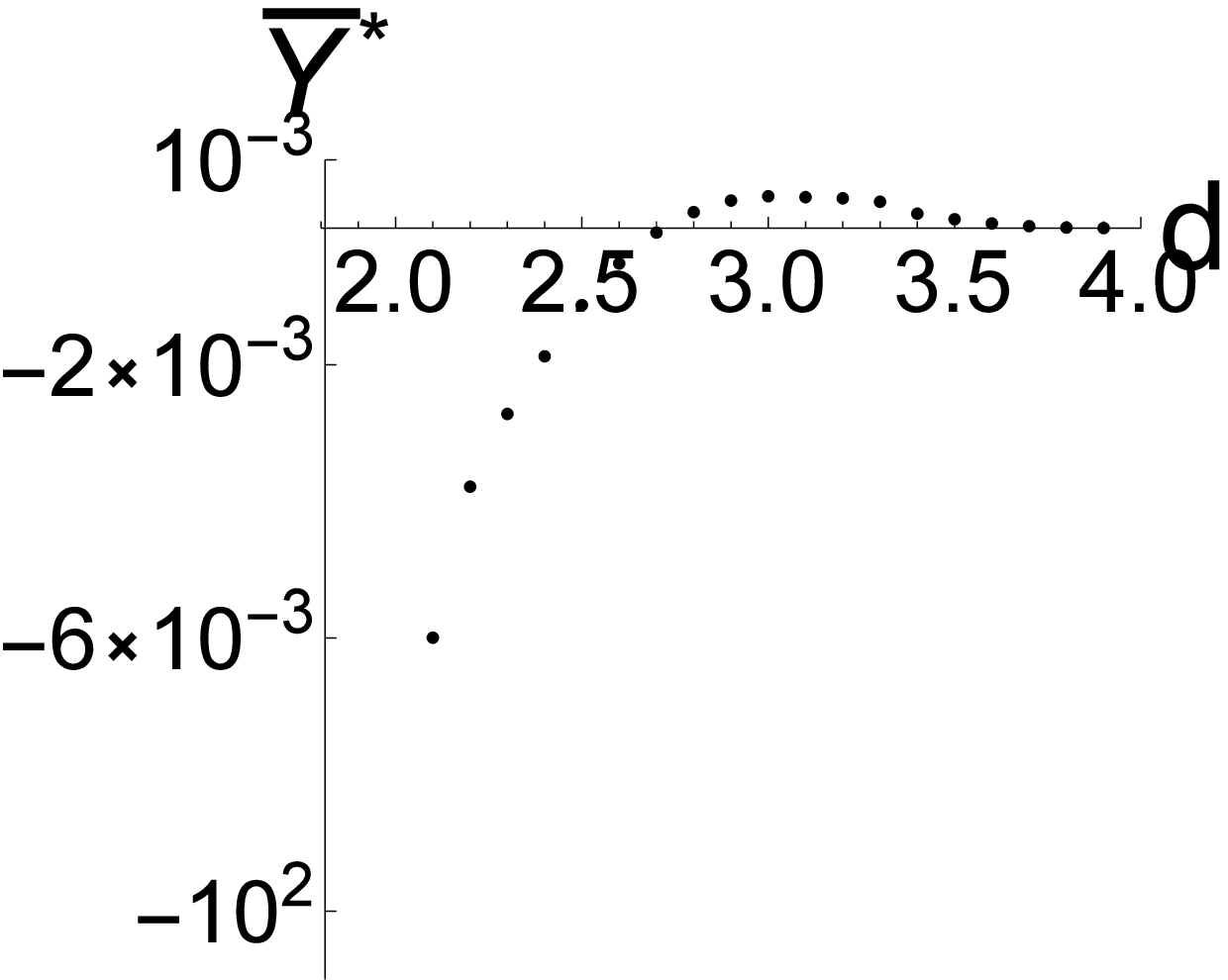,height=3.52cm,width=4.04cm,angle=0}
\psfig{file=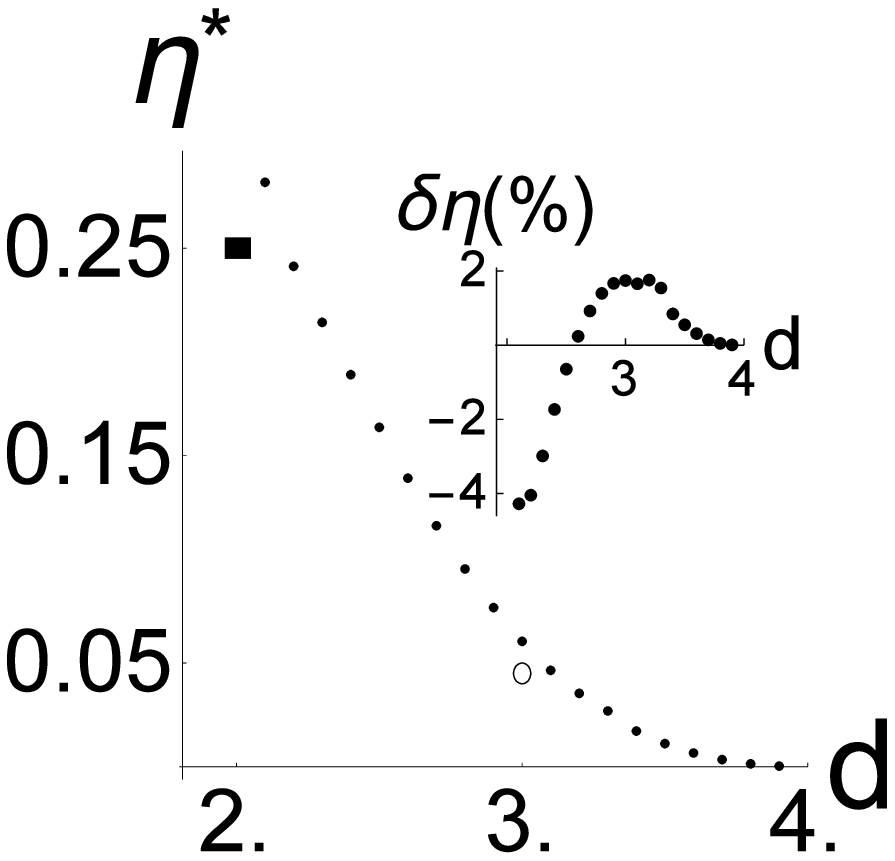,height=3.52cm,width=4.04cm,angle=0}}
\centerline{\psfig{file=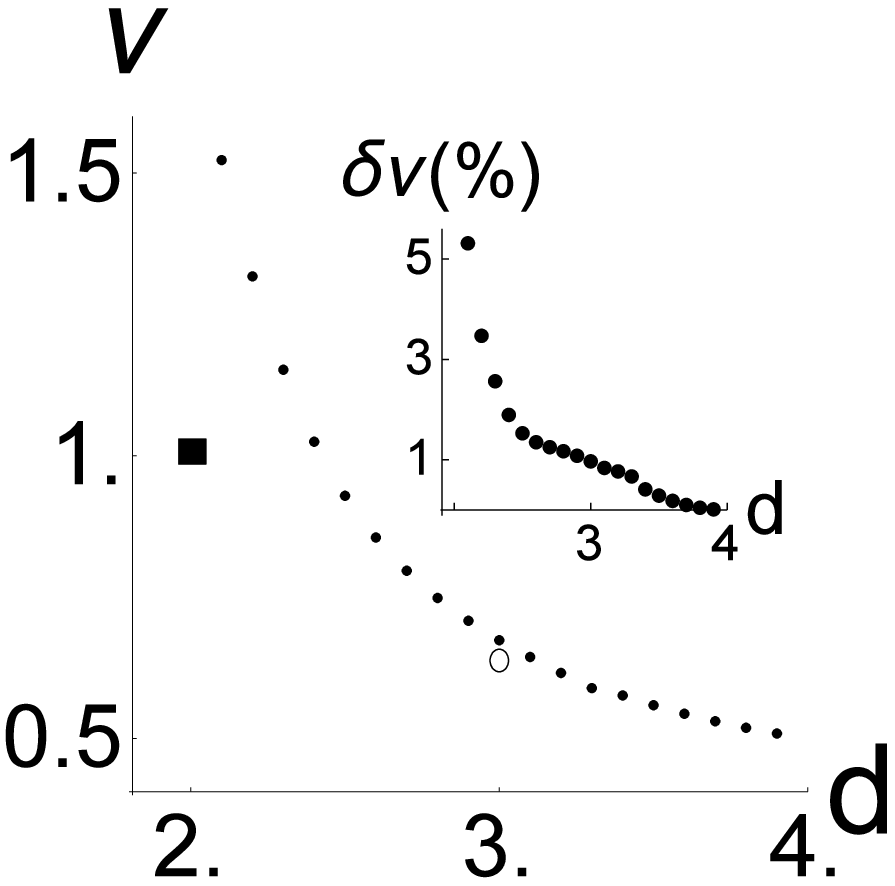,height=3.52cm,width=4.04cm,angle=0}}
\caption{\label{fig:ystdNNLO} Various quantities characterizing 
 the  WF FP  vs. the continuous dimension $d$ in the RG scheme NNLO$6$.
The values taken at $d=4$ are in agreement with those given by the $\epsilon$-expansion, Eq. \eq{4meps} for $\epsilon=0$. The insets show the NNLO effects
$\delta\eta= 100\times (\eta^{NNLO}-\eta^{NLO})/\eta^{NNLO}$ and 
$\delta \nu= 100\times (\nu^{NNLO}-\nu^{NLO})/\nu^{NNLO}$ on the anomalous dimension $\eta$ and the critical exponent $\nu$, respectively. We have also plotted
Onsager's exact results (full squares) and the NLOu values obtained in \cite{Tetra1993} (empty circles).
}
\end{figure}

We have also evaluated the dependence of the various quantities characterizing the WF FP as the function of the continuous dimension $d$ in the interval $2<d<4$ by solving the RG evolution equations numerically for the trajectories of bunch $\c{B}$ and identifying the WF crossover region. Here the correlation length's exponent $\nu$  was evaluated in a more straightforward and less time-consuming way from the derivative relation 
\bea\label{dernu}
 \nu^{-1}_\beta&=& - \frac{ \partial \beta_{\b{\kappa}} }{ \partial\b{\kappa}}\biggr|_{\b{\kappa}^*,\b{\lambda}^*,\b{Y}^*}
\eea
 like in \cite{Wette200310}. The results are shown in  Fig.  \ref{fig:ystdNNLO}.
The qualitative behaviour of the anomalous dimension  $\eta^*(d)$ and
that of the critical exponent $\nu(d)$ known from Ref. \cite{Wette200310} have been reproduced.
It was found that $\b{\kappa}^*(d)$, $\eta^*(d)$, and $\nu(d)$ fall off strictly monotonically with increasing dimension $d$ in the interval $d\in [ 2,4]$, while $\b{\lambda}^*(d)$ exhibits a maximum at $d\approx 3.5$.   The coupling $\b{Y}^*$ of the $\ord{\partial^4}$ term takes positive values for $d_0\approx 2.7 <d<4$ with a maximum at $d_{max}\approx  3.0$ and  exhibiting a zero at $d_0$.
 It decreases to negative values with decreasing dimension $d$ in the interval  $d_0>d>2$ indicating that the Euclidean action is not bounded from below for these dimensions in the RG scheme used.
 This may happen due to the neglection of the field-dependence of the derivative couplings or the truncation of the GE at terms of the order $\ord{\partial^4}$. In the latter case additional higher-derivative terms might turn the fixed-point action into a one bounded from below. In the insets of the plots of $\eta^*(d)$ and $\nu(d)$ in Fig. \ref{fig:ystdNNLO} we see  that the NNLO effect results in increments not exceeding $2$ per cents  of the values of $\eta$
and $\nu$  in the validity range $d_0<d<4$ of the applied RG scheme and it becomes more important with decreasing dimension $d$ in the interval $2\le d<d_0$.
 For $d=2$ our results deviate significantly from Onsager's exact values, but one has to keep in mind that $d=2$ lies out of the range of the validity of our RG scheme. We also see on the insets of  \ref{fig:ystdNNLO} that the NNLO effect on $\eta^*$  vanishes for  dimensions $d=4-\epsilon$ with $\epsilon \to 0^+$ as an effect of $\ord{\epsilon^3}$  on the quantity of $\ord{\epsilon^2}$.
Although being  small the NNLO effects on $\eta^*$ and $\nu$ may be
 comparable to the effect of the inclusion of the field-dependence of the wavefunction renormalization (see e.g. \cite{Canet2003,Wette200310}).

\section{$O(N)$ models for $N\ge 2$ and $d=3$}\label{onmods}

\subsection{Evolution equations}
For the $O(N)$ models with $N\ge 2$ the dimensionless couplings are defined like in the $O(1)$ case except that the
dimensionless field-variable $\b{r}$ is defined by the help of the field-independent wavefunction renormalization of the Goldstone-modes as $\b{r}=Z_{\perp k} k^{-(d-2)}r$. The explicit forms of the evolution equations  for the dimensionless
 couplings have been generated in the NLO and NNLO approximations for various truncations up to  $M=6$ from Eqs. \eq{evU}-\eq{cZperev}  by means of computer algebra.

It is instructive to write out   the evolution equations in the NLO$2$ scheme,
\bea\label{bkaN}
\beta_{\b{\kappa}}&=& -(d-2+\eta)\b{\kappa} +\frac{2\alpha_d}{d}(3\h{z}\b{g}^2
  +N-1),\\
\label{blaN}
\beta_{\b{\lambda}} &=& (d-4+2\eta)\b{\lambda}+\frac{2\alpha_d}{d}\b{\lambda}^2
[ 9\h{z}\b{g}^3+ 2(N-1)] ,\\
\label{bhzN}
\beta_{\h{z}}&=&-\h{z}(\b{\eta}-\eta),
\eea
where we have introduced the propagator
\bea
 \b{g}&=& (\h{z} +2\b{\kappa}\b{\lambda})^{-1},
\eea
the ratio
\bea\label{hz}
  \h{z}&=& Z_{\perp k}^{-1}Z_{\parallel k}, 
\eea
and the anomalous dimensions via the relations
\bea
  {\dot Z}_{\parallel k}&=&-\b{\eta} Z_{\parallel k},\nn
 {\dot Z}_{\perp k}&=&-\eta Z_{\perp k},
\eea
where
\bea\label{andb}
  \b{\eta}&=& \frac{4\alpha_d}{d}\b{\kappa}\b{\lambda}^2\biggl\lbrack
  \frac{d-2}{d+2} \frac{N-1}{\h{z}}
 +9\h{z}\b{g}^4 \biggl( 1-\frac{4\h{z}\b{g} }{d+2}\biggr) \biggr\rbrack,\\
\label{and}
{\eta}&=&  \frac{16\alpha_d}{d(d+2)}\b{\kappa}\b{\lambda}^2\h{z}\b{g}^2
[ d -2\h{z} \b{g} ].
\eea
We see that assuming uniform wavefunction renormalization, one can work with the single coupling $\h{z}$ instead of the wavefunction renormalizations for the radial and the Goldstone modes of the field. This holds for the NNLO approximation too. 

\subsection{Numerical results}

Our numerical investigation of the $O(N)$ models has been restricted to the number of dimensions $d=3$. The RG evolution has always been started from an $O(N)$ symmetric state with $Z_{\pa\Lambda}=Z_{\pe\Lambda}=1$ ($\h{z}(\Lambda)=1$) and
with vanishing higher-derivative terms, $\b{Y}_{\pa\Lambda}=\b{Y}_{\pe\Lambda}=0$. 
The numerical search for the WF FP and its characteristics has been performed applying two different procedures:  Procedure A and  Procedure B.  Both procedures consisted of two steps. At first,  we solved the  RG evolution equations for the bunch $\c{B}$ of the trajectories and  identified the crossover region. At second, the fixed-point equations were solved by means of a Newton-Rhapson algorithm started at the educated guess of the roots read off from the behaviour of the crossover region in the first step. In Procedure A  {\em (i)} the $O(N)$ symmetry of the EEA has been enforced on the NLO level by considering $\h{z}=1$ at all scales,
i.e., $Z_{\pa k}=Z_{\pe k}$ during the evolution and the FP was searched for setting  $\h{z}^*=1$; and {\em (ii)}   the  evolution equations for $\b{\kappa},~\b{\lambda},~\b{Y}_{\pa k},~\b{Y}_{\pe k}$ and the fixed-point equations $\beta_{\b{\kappa}}=\beta_{\b{\lambda}}=\beta_{\b{Y}_\pa}=\beta_{\b{Y}_\pe}=0$ have been solved by making use of the explicit expressions for $\eta$ and $\b{\eta}$.  The critical exponent $\nu$ has also been determined for various $N$ values by making use of the same numerical procedure as that applied in the case of the 3-dimensional $O(1)$ model. The critical exponent $\nu$ has also been determined like in the case of the $O(1)$ model,   by investigating the `temperature-dependence'
of the correlation length $\xi\sim 1/k_c$ at the lower end of the crossover region.
In Procedure  B {\em (i)} the ratio  $\h{z}$ has also been evolved, i.e.,  the evolution Eq. \eq{bhzN} has been included into the set of the evolution equations; and {\em (ii)}   the fixed-point equations  $\beta_{\b{\kappa}}=\beta_{\b{\lambda}}=\beta_{\b{Y}_\pa}=\beta_{\b{Y}_\pe}=0$ have been solved for setting  $\h{z}^*=\h{z}(k_c)$ determined in the first step of the procedure at the scale $k_c$ for which it holds $\b{\kappa}(k_c)=0$. The reason that we performed our investigation
of the NNLO effect by means of both procedures is that setting $\h{z}=1$ fails to solve the equation $\beta_{\h{z}}=0$ even in the NLO approximation, but one finds at the WF FP $\beta_{\h{z}}\sim -\ord{\b{\eta}}$.

\begin{figure}[htb]
\centerline{\psfig{file=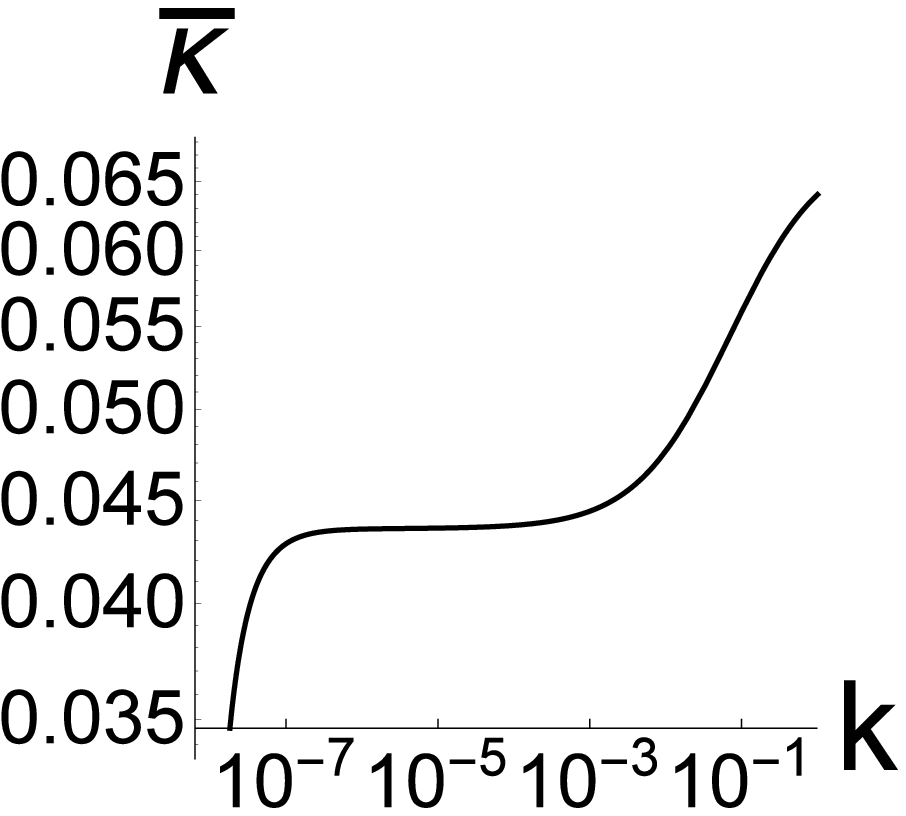,height=3.52cm,width=4.04cm,angle=0}
\psfig{file=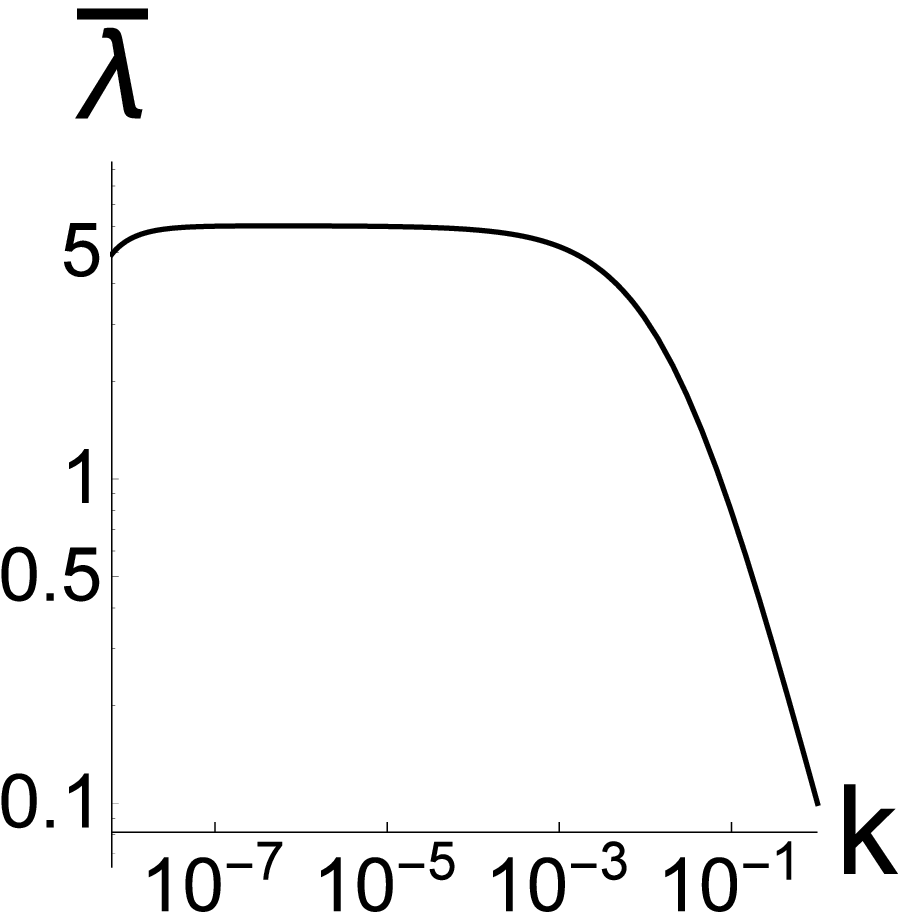,height=3.52cm,width=4.04cm,angle=0} }
\centerline{
\psfig{file=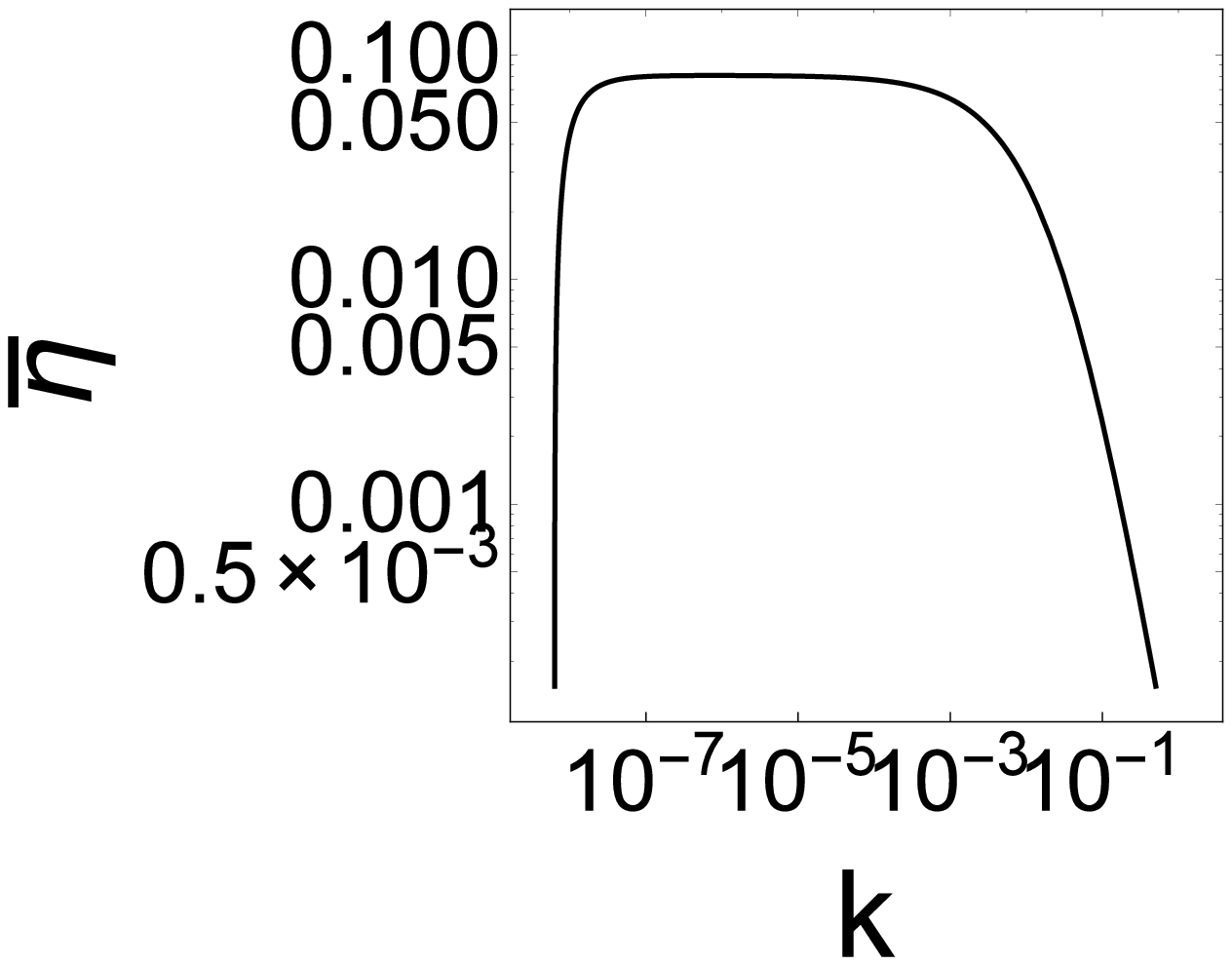,height=3.52cm,width=4.04cm,angle=0}
\psfig{file=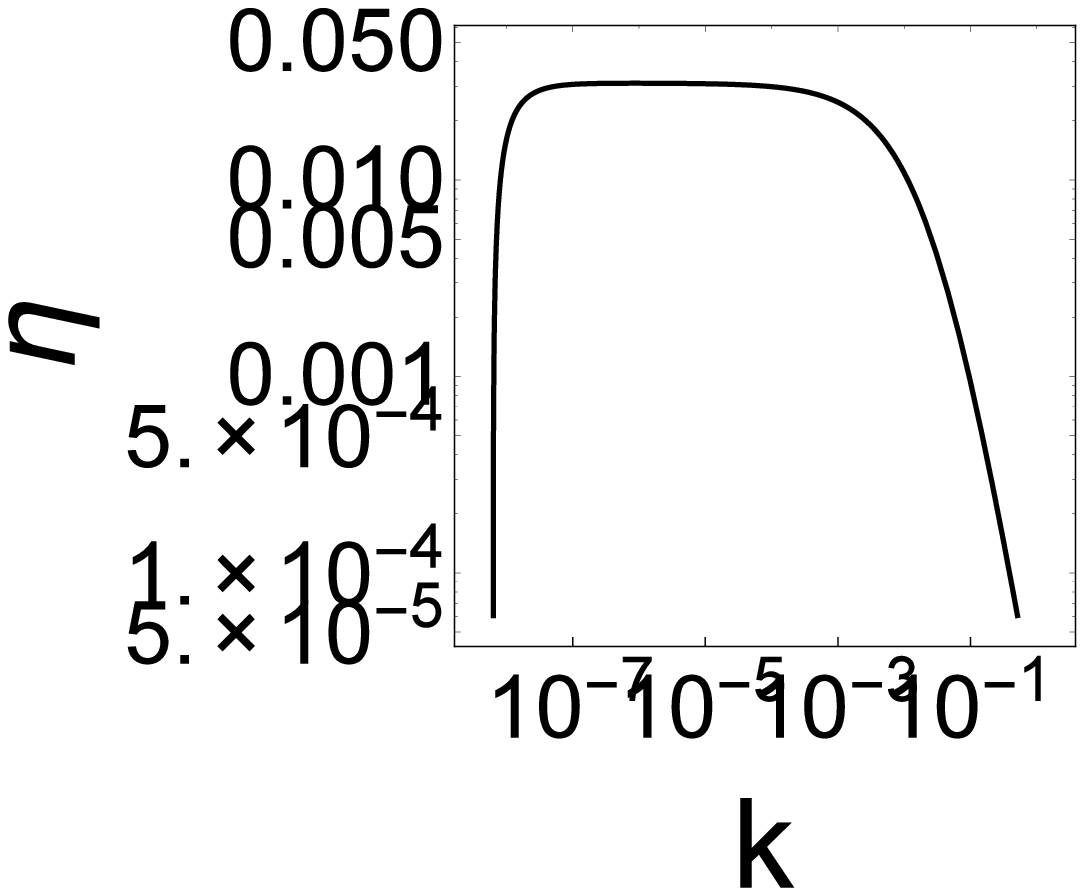,height=3.52cm,width=4.04cm,angle=0}
}
\caption{\label{fig:nnlo6_o2_z1} Scale-dependences of the couplings
$\b{\kappa}$ and $\b{\lambda}$ as well as those of the anomalous dimensions $\b{\eta}$ and $\eta$ of the radial and the Goldstone modes, respectively, 
on the nearly critical trajectories of the $O(2)$ model for dimension $d=3$ in NNLO6 approximation
for $\h{z}(k)=1$.  }
\end{figure}

\begin{figure}[htb]
\centerline{
\psfig{file=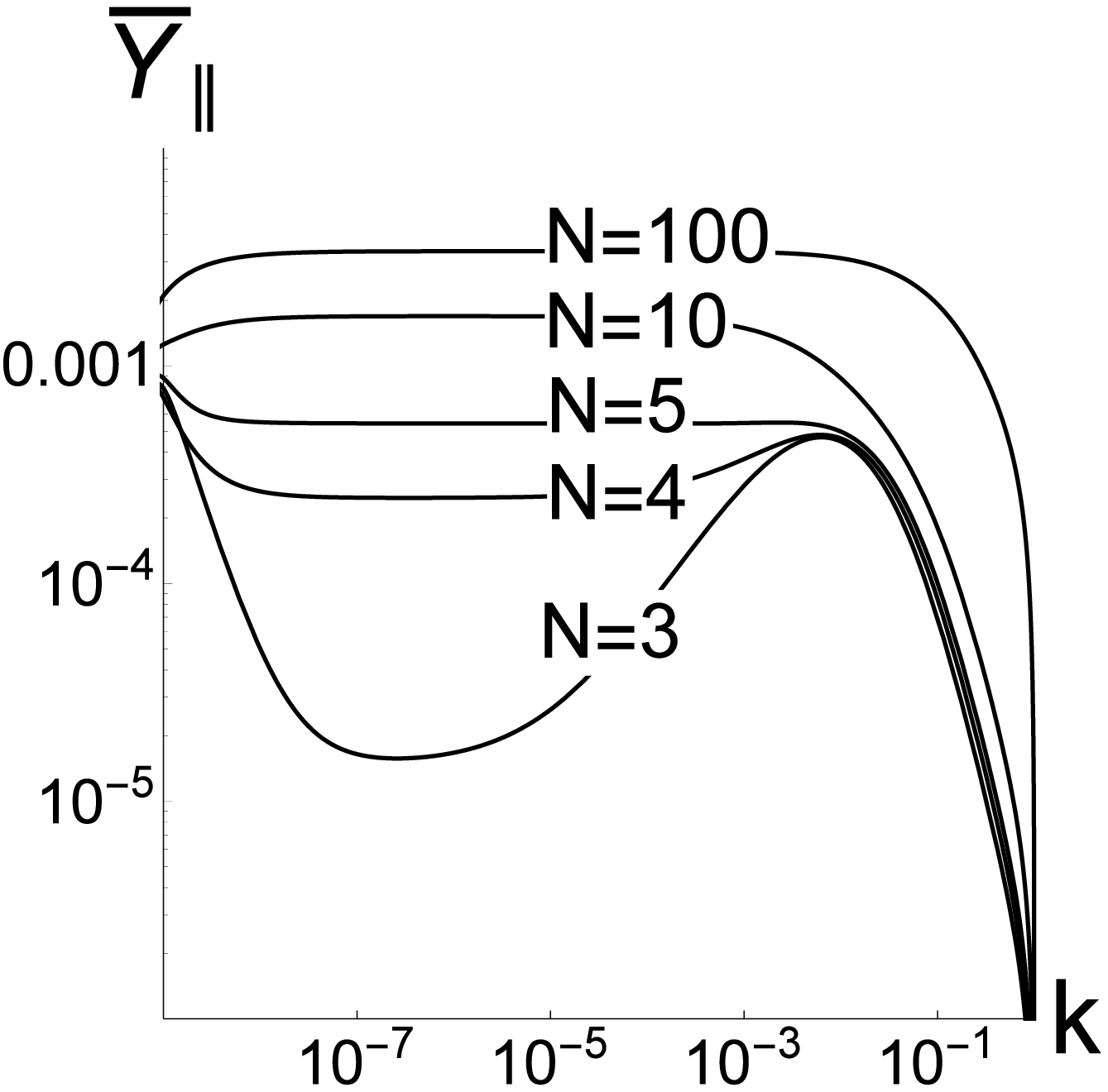,height=3.52cm,width=4.04cm,angle=0}
\psfig{file=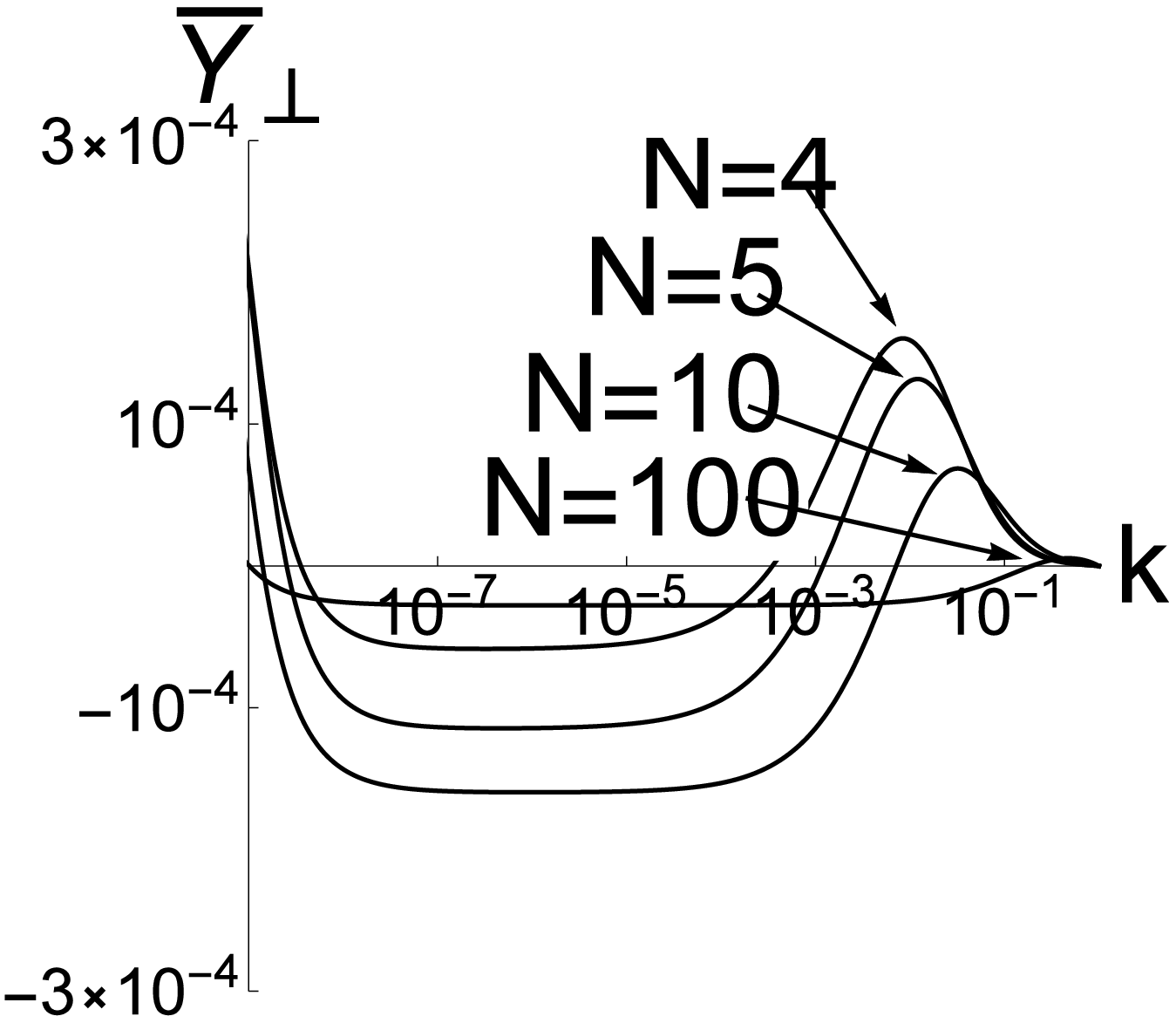,height=3.52cm,width=4.04cm,angle=0}
}
\caption{\label{fig:nnlo6_onycomp} 
Scaling of the higher-derivative couplings  $\b{Y}_\pa$ and $\b{Y}_\pe$ for the radial and the Goldstone modes, respectively, for various values of $N$ in the NNLO6 approximation  evaluated with $\h{z}=1$.}
\end{figure}

\begin{figure}[htb]
\centerline{
\psfig{file=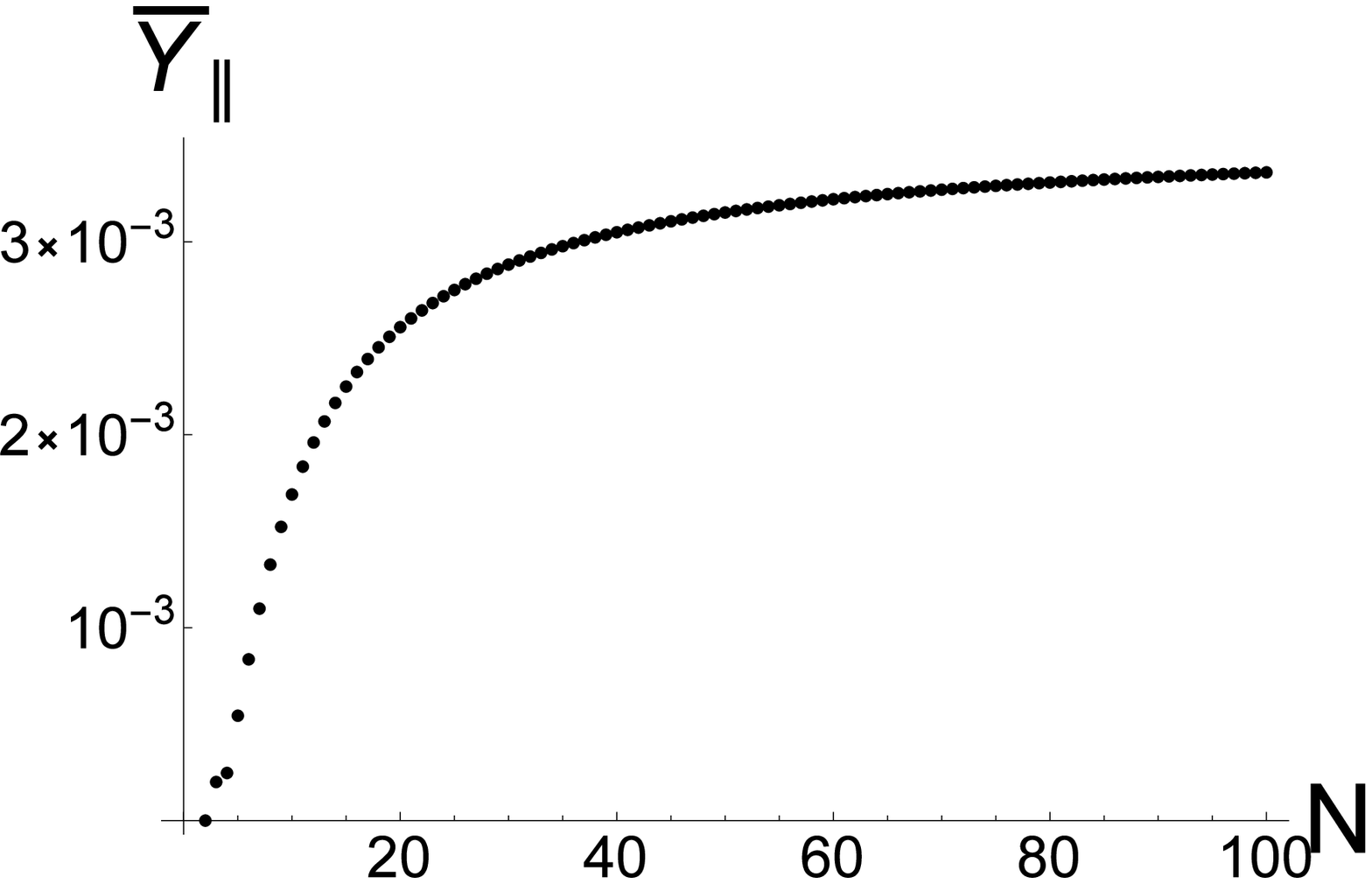,height=3.52cm,width=4.04cm,angle=0}
\psfig{file=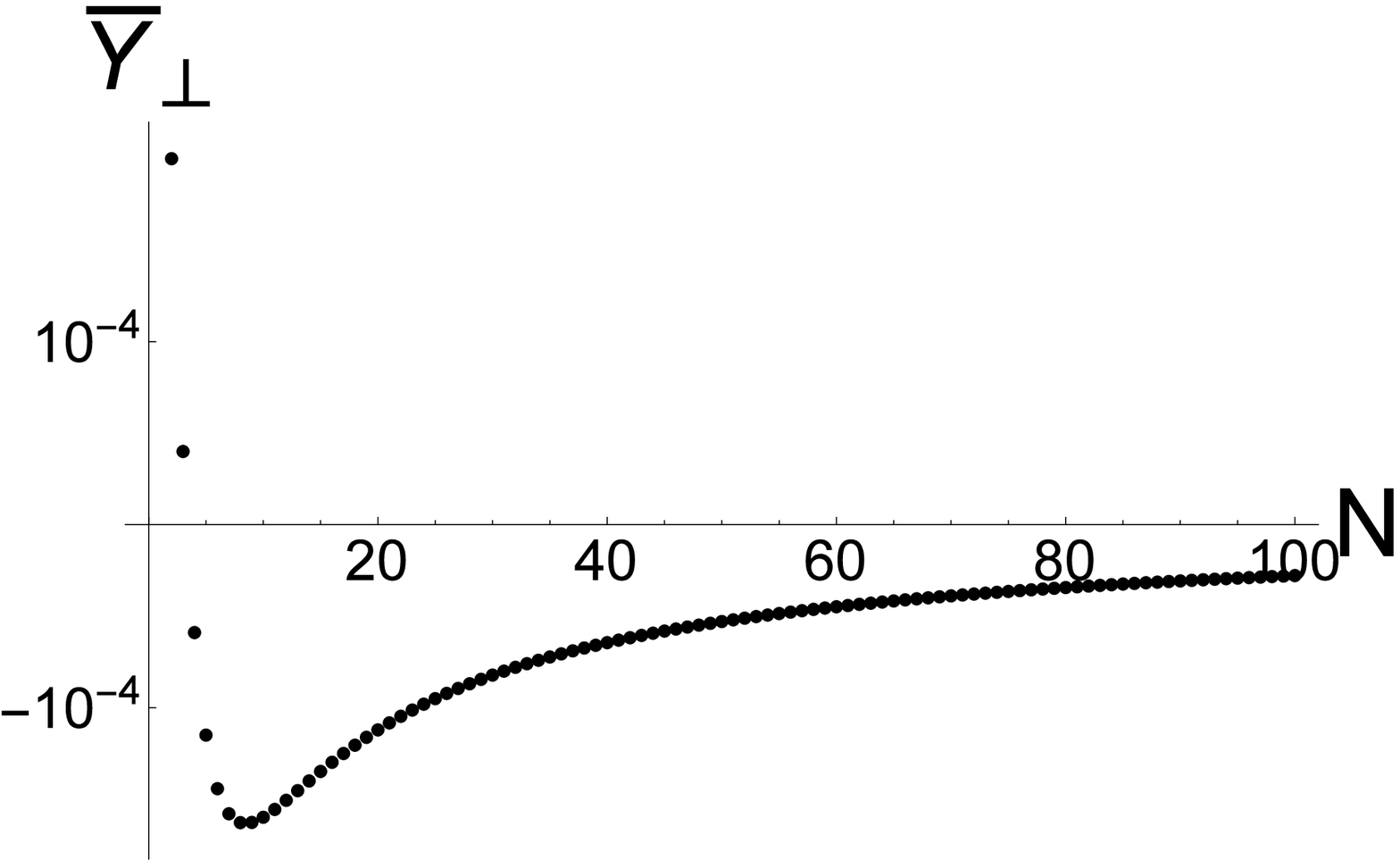,height=3.52cm,width=4.04cm,angle=0}
}
\caption{\label{fig:nnlo6_ony} $N$-dependences  of the 'plateau' values of the
higher-derivative couplings  $\b{Y}_\pa$ and $\b{Y}_\pe$ for the radial and the Goldstone modes, respectively, evaluated with $\h{z}=1$.
}
\end{figure}

\begin{figure}[htb]
\centerline{
\psfig{file=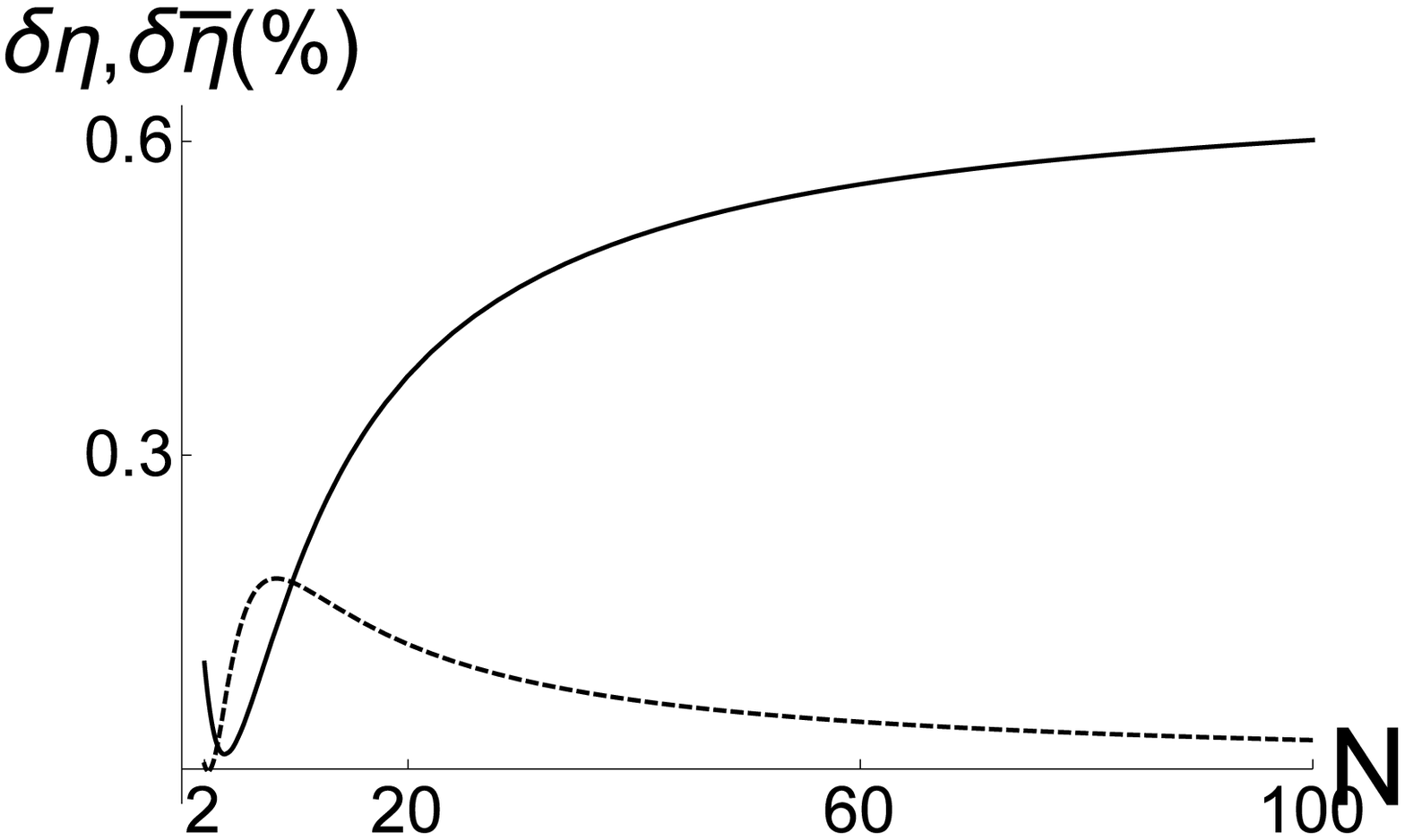,height=3.52cm,width=4.04cm,angle=0}
\psfig{file=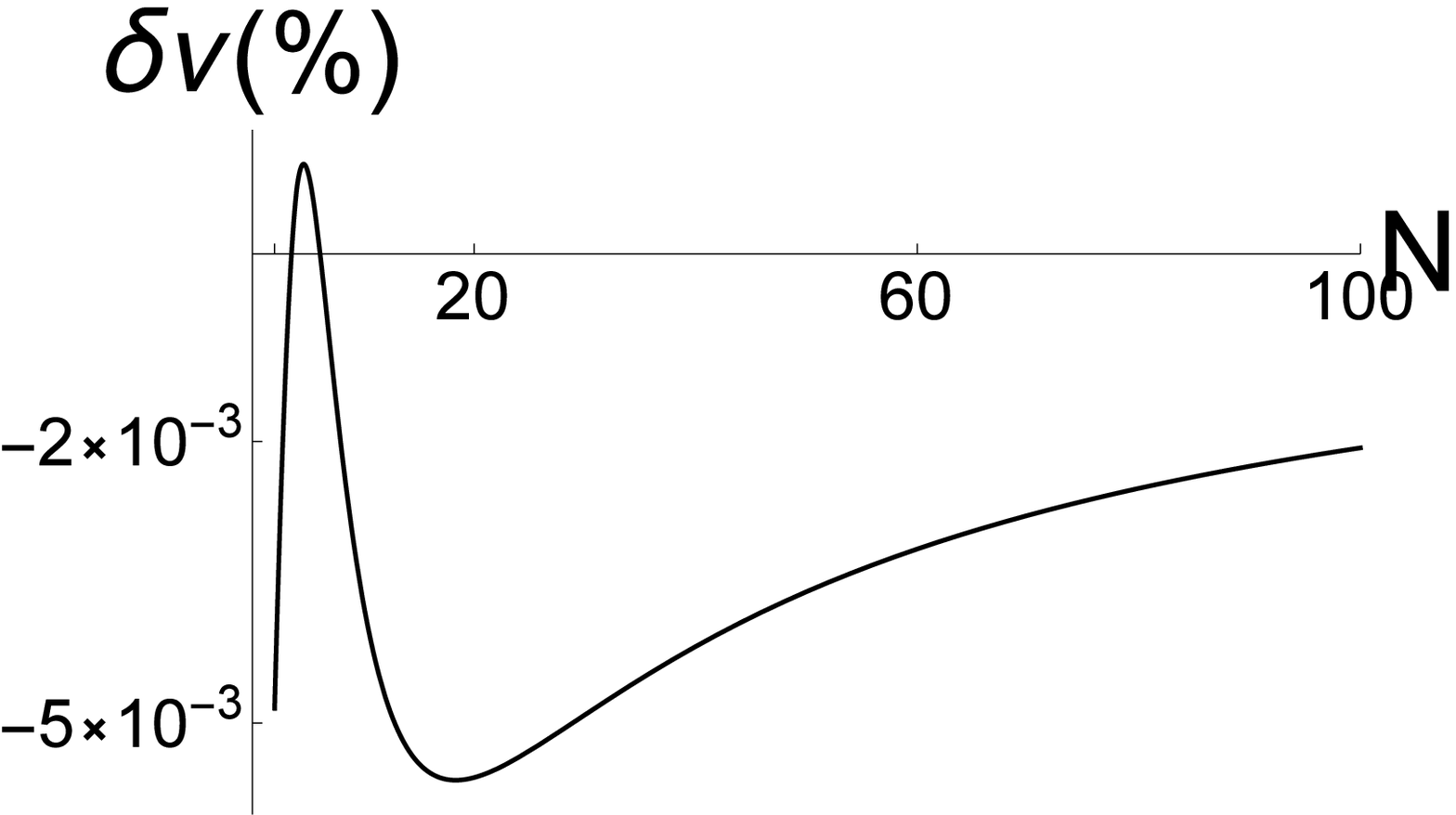,height=3.52cm,width=4.04cm,angle=0}
}
\caption{\label{fig:on_gecomp} $N$-dependence of the NNLO effect on the various critical exponents $\b{\eta}^*$, $\eta^*$, and $\nu$  in the approximation  NNLO$6$, evaluated with $\h{z}=1$. Here the NNLO effect in per cents is given via $\delta f= 100\times (f^{NNLO}-f^{NLO})/f^{NNLO}$ for $f=\b{\eta}$ (solid line), $f=\eta$ (dashed line) in the plot to the left, and $f= \nu$ in the plot to the right.
}
\end{figure}

The results of  Procedure A are shown in Figs. \ref{fig:nnlo6_o2_z1},  \ref{fig:nnlo6_onycomp},  \ref{fig:nnlo6_ony}, and \ref{fig:on_gecomp} and the numerical values of the characteristics of the WF FP obtained in  the NNLO6 approximation are listed in Table
\ref{tab:oncritexp_NNLO6}.

\begin{table}[htb]
\begin{center}
\begin{tabular}{|c|c|c|c|c|c|c|c|c|}
\hline
 $N$       & $\b{\kappa}^*$ & ${\b{\lambda}}^*$   & $10^4\times \b{Y}_\pa^*$      & $10^4\times\b{Y}_\pe^*$  & $\b{\eta}^* $ & $\eta^*$    & $\nu$   \cr 
\hline\hline
 $1$       & $0.031$    & $6.02$        &$5.0$   & $-$                & $0.059$  &  $-$    & $0.634$    \cr
 $2$       & $0.043$    & $6.03$        &$\approx 0.0$ &$2.0$  & $0.077$  &  $0.0320$  & $0.700$ \cr
 $3$       & $0.057$    & $5.35$        &$2.0$   &$0.4$  & $0.085$  &  $0.0300$  & $0.739$ \cr
$4$        & $0.072$    &  $4.73$       &$2.4$  & $-0.6$  & $0.088$  &
 $0.028$& $0.775$ \cr
 $5$       & $0.087$    &  $4.19$       &$5.5$  & $-1.0$  & $0.090$  &
 $0.026$& $0.806$ \cr
 $10$      & $0.168$    & $2.54$        &$16.0$   &$-1.6$   & $0.095$  &  $0.0165$  & $0.896$  \cr
 $100$     & $1.68$     & $0.29$   &$33.0$    &$-0.3$     & $0.099$    &  $0.0020 $  & $0.990$ \cr
\hline
\end{tabular}
\end{center}
\caption{\label{tab:oncritexp_NNLO6}  Position of the WF FP,  the critical exponent $\nu$, and the  anomalous dimensions $\b{\eta}^*$ and $\eta^*$ for the radial and the Goldstone modes, respectively, for various values of $N$  in  NNLO6  approximation evaluated with $\h{z}=1$.}
\end{table}

The couplings $\b{\kappa}$, $\b{\lambda}$ and the anomalous dimensions $\b{\eta}$ and $\eta$ show up quite similar scale-dependences for $2\le N\le 100$ (see Fig. \ref{fig:nnlo6_o2_z1} for typical scalings). Namely, they exhibit a  cca. four-order-of-magnitude wide WF crossover region   in which they keep their constant values. These 'plateau' values match almost perfectly (up to $3$ to $4$ valuable digits)  the corresponding   values obtained by solving the fixed-point equations and given in Table  \ref{tab:oncritexp_NNLO6}. Similarly to the case  $O(1)$, beyond the lower end  of the WF crossover region the parameter $\b{\kappa}$ as well as the anomalous dimensions $\b{\eta}$ and $\eta$ fall suddenly to zero.
 Similarly to the other running couplings, the scale-dependences of the  higher-derivative couplings $\b{Y}_\pa$ and $\b{Y}_\pe$ for the radial and the Goldstone modes exhibit  crossover regions at around $k\sim 10^{-6}$  with constant values  (see Fig. \ref{fig:nnlo6_onycomp}) and the latter are in good agreement again with the values obtained by solving the fixed-point equations. The 'plateaus' of the higher-derivative couplings broaden with increasing $ N$. As shown in Fig.  \ref{fig:nnlo6_ony} the `plateau' values of $\b{Y}_\pa$ increase monotonically and saturate, while those of $\b{Y}_\pe$ have a minimum at around $N\approx 10$ and saturate for large $ N$ too.
It was found that the higher-derivative coupling  $\b{Y}_\pe$ of the Goldstone modes   exhibit negative  'plateau' values  for $N>3$ (see Figs. \ref{fig:nnlo6_onycomp} and \ref{fig:nnlo6_ony}). Therefore the approximations used lead to a critical theory with action unbounded from below for $N>3$. This  may happen due to either the restriction to field-independent derivative couplings or the lack of even higher derivative terms in the EAA. The clarification of this problem needs further investigations in more sophisticated RG frameworks which is out of the scope of the present paper.

As shown in Fig. \ref{fig:on_gecomp}, the NNLO effect, i.e., the inclusion of
 the running higher-derivative terms generally results in  increments of the values of  the anomalous dimensions $\b{\eta}^*$, $\eta^*$, and a decrement of the critical exponent $\nu$, as a rule. The NNLO effect $\delta \eta$ on the anomalous dimension of the radial mode saturates for asymptotically large $N$ values, but does not exceed 1 per cent, while the NNLO effects $\delta \b{\eta}$ on the anomalous dimension of the Goldstone modes dies out for large $N$ values and never exceeds $\sim 0.2$ per cents. The NNLO effect $\delta \nu$ on the correlation length's critical exponent is very small and dies out too with
asymptotically increasing $N$. 

Our results obtained in Procedure A by means of the ansatz 
\eq{eaaon} (with the choice in Eq. \eq{invpro}) show up the rather striking feature that the critical values of $\b{\eta}^*$ and $\eta^*$ turned out  to be quite different, even in the  NLO approximation, 
 although setting  $\h{z}=1$ should enforce by definition that the wavefunction renormalizations for the radial and the Goldstone modes were identical at all scales. Even more the higher-derivative couplings $\b{Y}_\pa$ and $\b{Y}_\pe$ for the radial and the Goldstone modes are far from being identical, although there critical values remain  cca. 4 orders of magnitude smaller than the value $\h{z}=1$.   The point is that the fixed-point equation $\beta_{\h{z}}=0$ is not satisfied for $\h{z}=1$ when the 'plateau' values $\b{\kappa}^*,~ \b{\lambda}^*,~ \b{Y}^{\pe *},~\b{Y}^{\pa *}$ are inserted in the explicit formulas for $\b{\eta}$ and $\eta$ (like those in Eqs.
\eq{andb} and \eq{and}), the right-hand side of Eq. \eq{bhzN} takes the value
$\approx -\b{\eta}^*$. Therefore, in the RG framework used there occurs an inconsistency  of the order of the anomalous dimension of the radial mode.    
Thus one has to conclude that in the applied RG framework one obtains a critical theory at the WF FP in which a slight  explicit  symmetry breaking is  present in the gradient terms. Let us put it in another way: the critical theory preserves $O(N)$ symmetry with an accuracy of the order of the anomalous dimension $\b{\eta}^*$.

\begin{table}[htb]
\begin{center}
\begin{tabular}{|c||c|c||c|c|}
\hline
$N$     & $\nu$  & $\nu$ \cite{Tetra1993} & $\eta^*$  & $\eta$ \cite{Tetra1993}  \cr
\hline\hline
$1$  & $0.634$ & $0.638$  & $0.059$  & $0.045$  \cr
$2$  & $0.700$ & $0.700$  & $0.032$   & $0.042$  \cr
$3$  & $0.739$ & $0.752$  & $0.030$  & $0.038$  \cr
$4$  & $0.775$ & $0.791$  & $0.028$  & $0.034$  \cr
$10$ & $0.896$ & $0.906$  & $0.0165$ & $0.019$  \cr
$100$& $0.990$ & $0.992$  & $0.002$  & $0.002$  \cr
\hline
\end{tabular}
\end{center}
\caption{\label{tab:oncomp} Comparison of our NNLO6u results (left columns)
 with the NLOu results of Wetterich's group (right columns) \cite{Tetra1993},
the latter obtained by the usage of the exponential regulator.
 }
\end{table}

Setting $\h{z}=1$ corresponds to an $O(N)$ symmetric ansatz
for the EAA at the NLO level and enables one to make comparisons with the NLOu results in Ref. \cite{Tetra1993} (see Table \ref{tab:oncomp}). This comparison shows that (i) the anomalous dimension determined in Ref.  \cite{Tetra1993} is essentially the anomalous dimension for the Goldstone modes, and (ii) our NNLOu data show the same qualitative dependences on $N$ as the NLOu data in \cite{Tetra1993} do. For the critical exponent $\nu$ of the correlation length the discrepancy is less than 2 per cents,
for the anomalous dimension $\eta^*$ of the Goldstone modes it makes 20 - 30 per cents, but seems to disappear for large $N$.  These discrepancies are much larger than the NNLO effect (see Fig. \ref{fig:on_gecomp}), consequently, they occur basicly at the NLO level and are caused essentially by  the usage of different regulators.

\begin{figure}[ht]
\centerline{\psfig{file=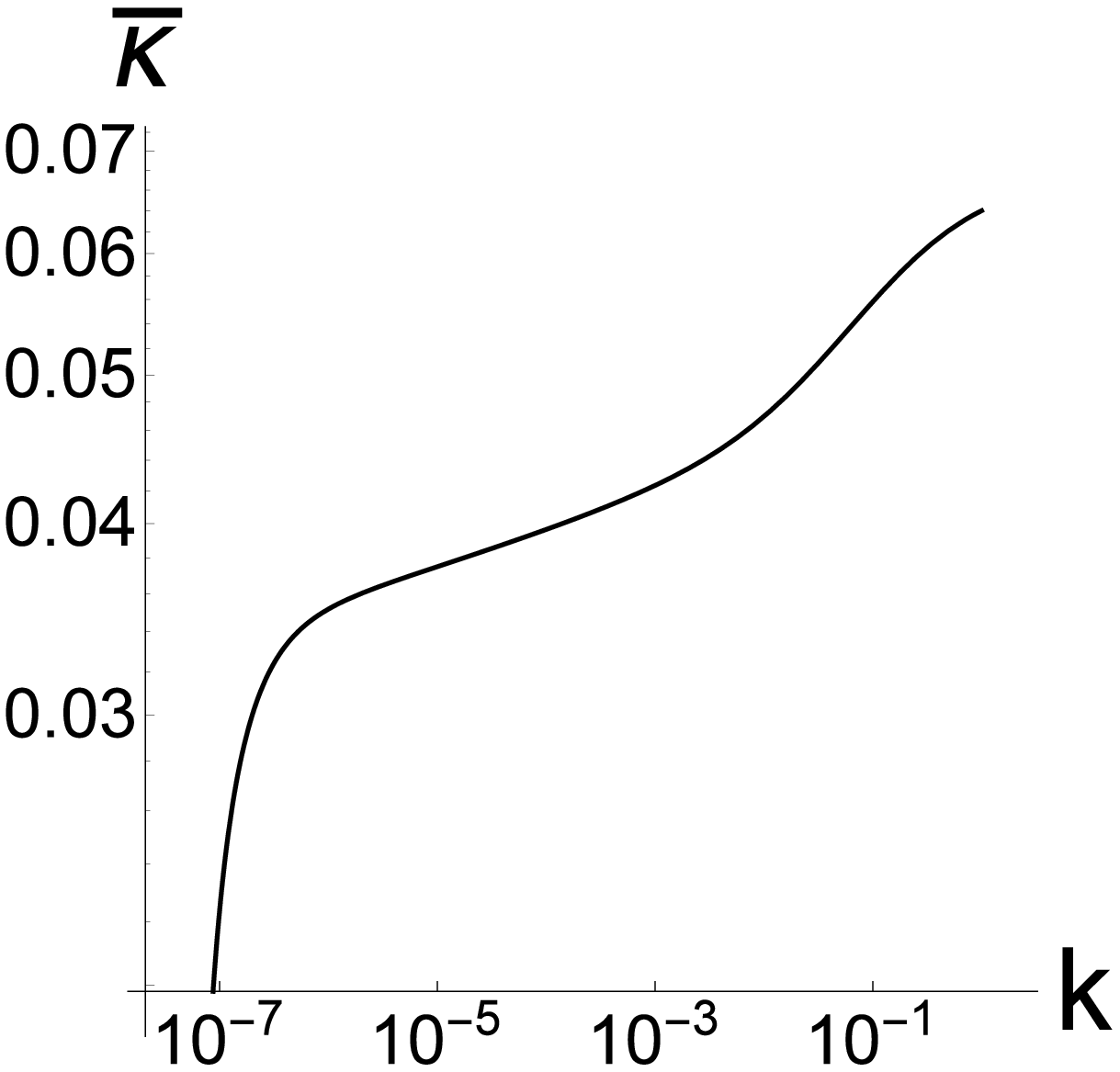,height=3.52cm,width=4.04cm,angle=0}
\psfig{file=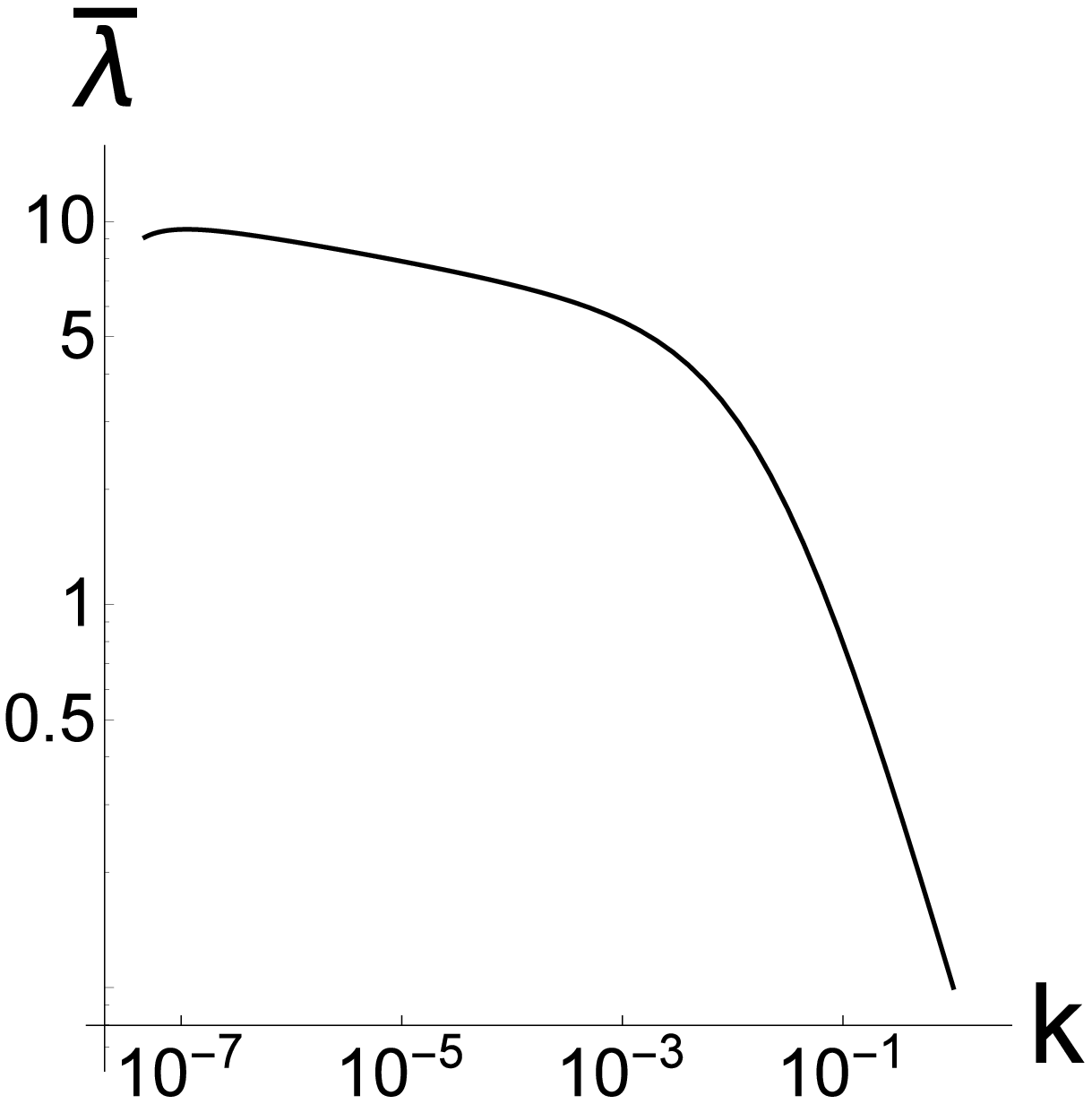,height=3.52cm,width=4.04cm,angle=0} }
\centerline{
\psfig{file=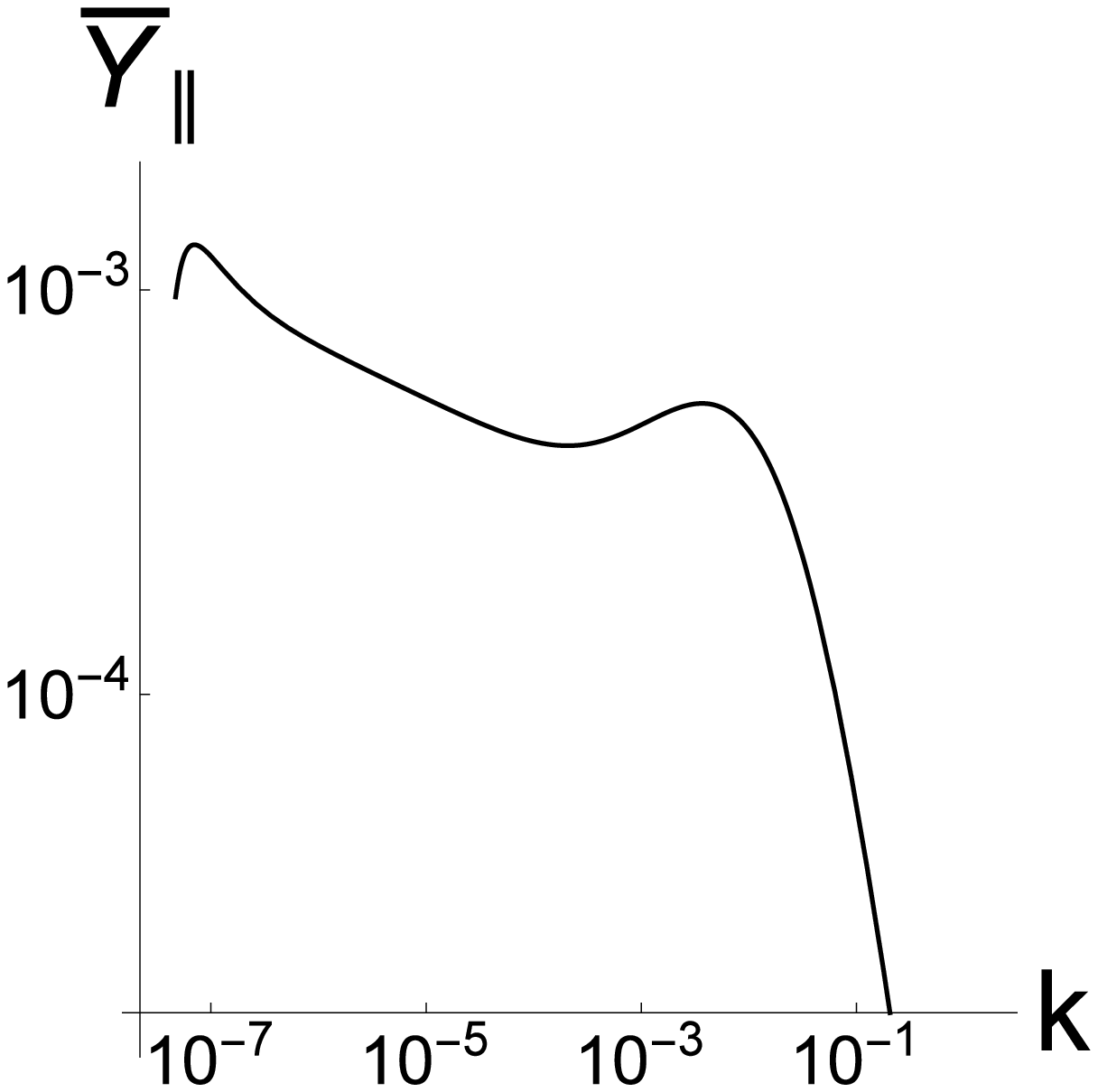,height=3.52cm,width=4.04cm,angle=0}
\psfig{file=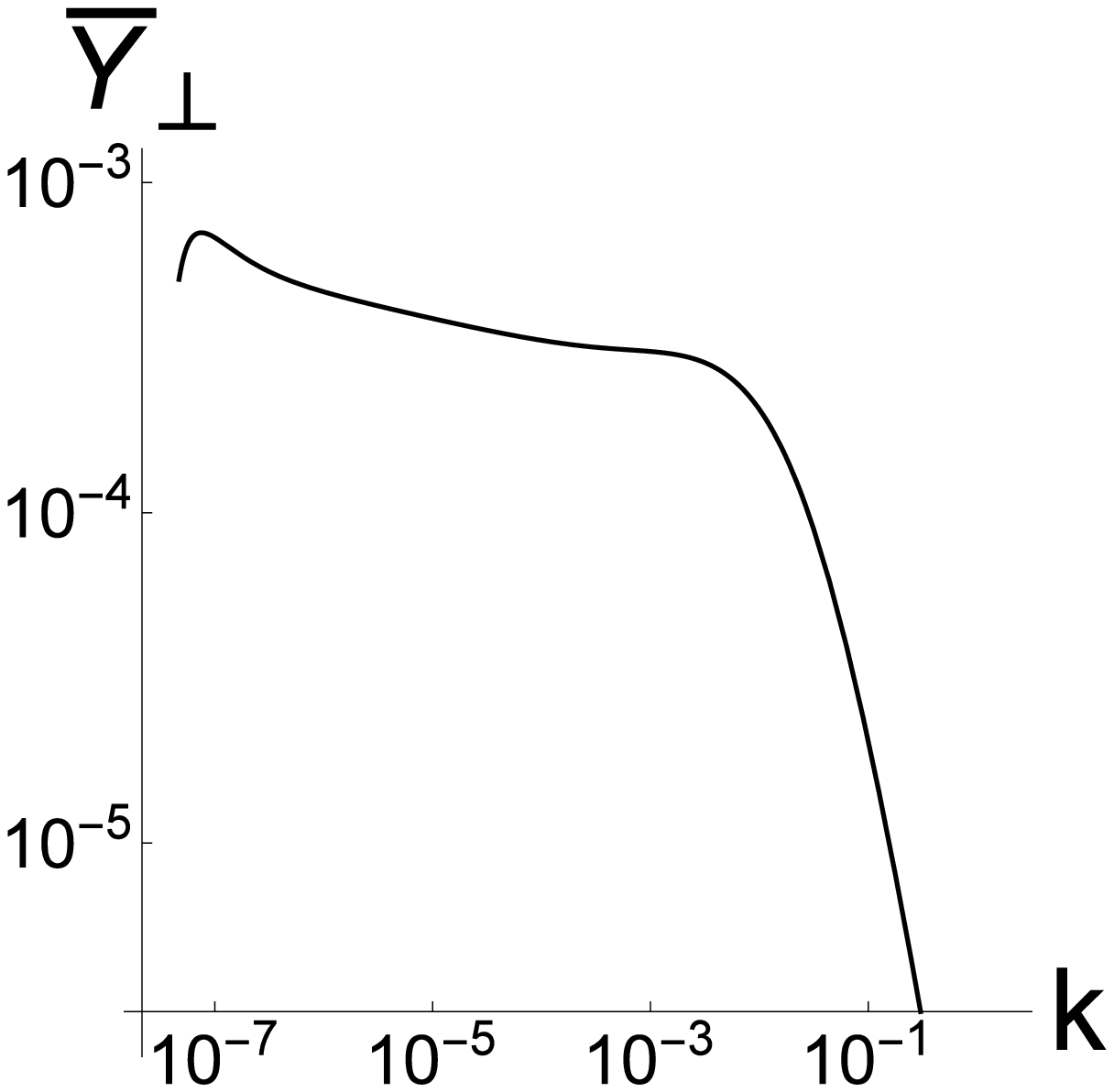,height=3.52cm,width=4.04cm,angle=0}
}
\centerline{\psfig{file=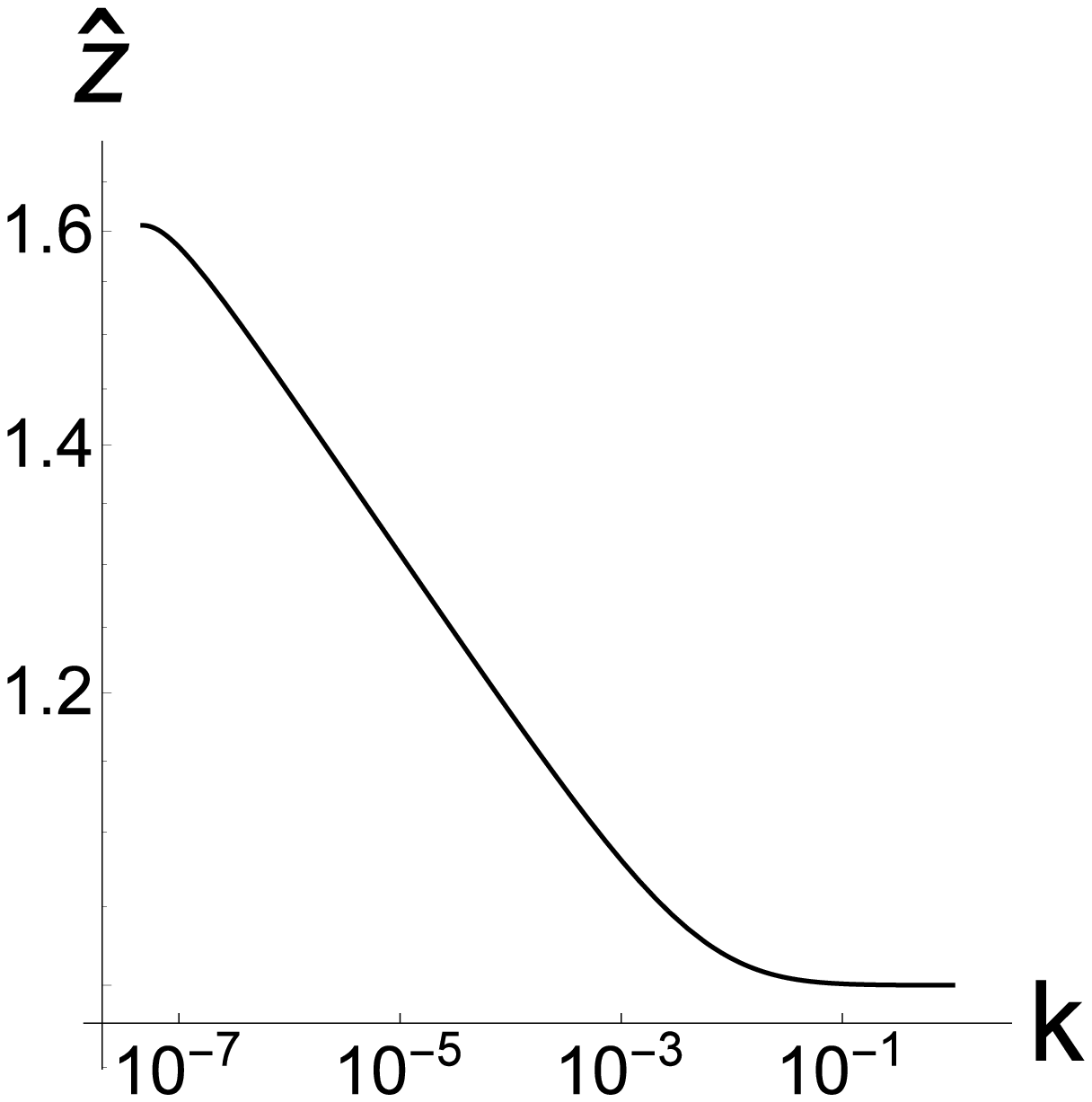,height=3.52cm,width=4.04cm,angle=0}
\psfig{file=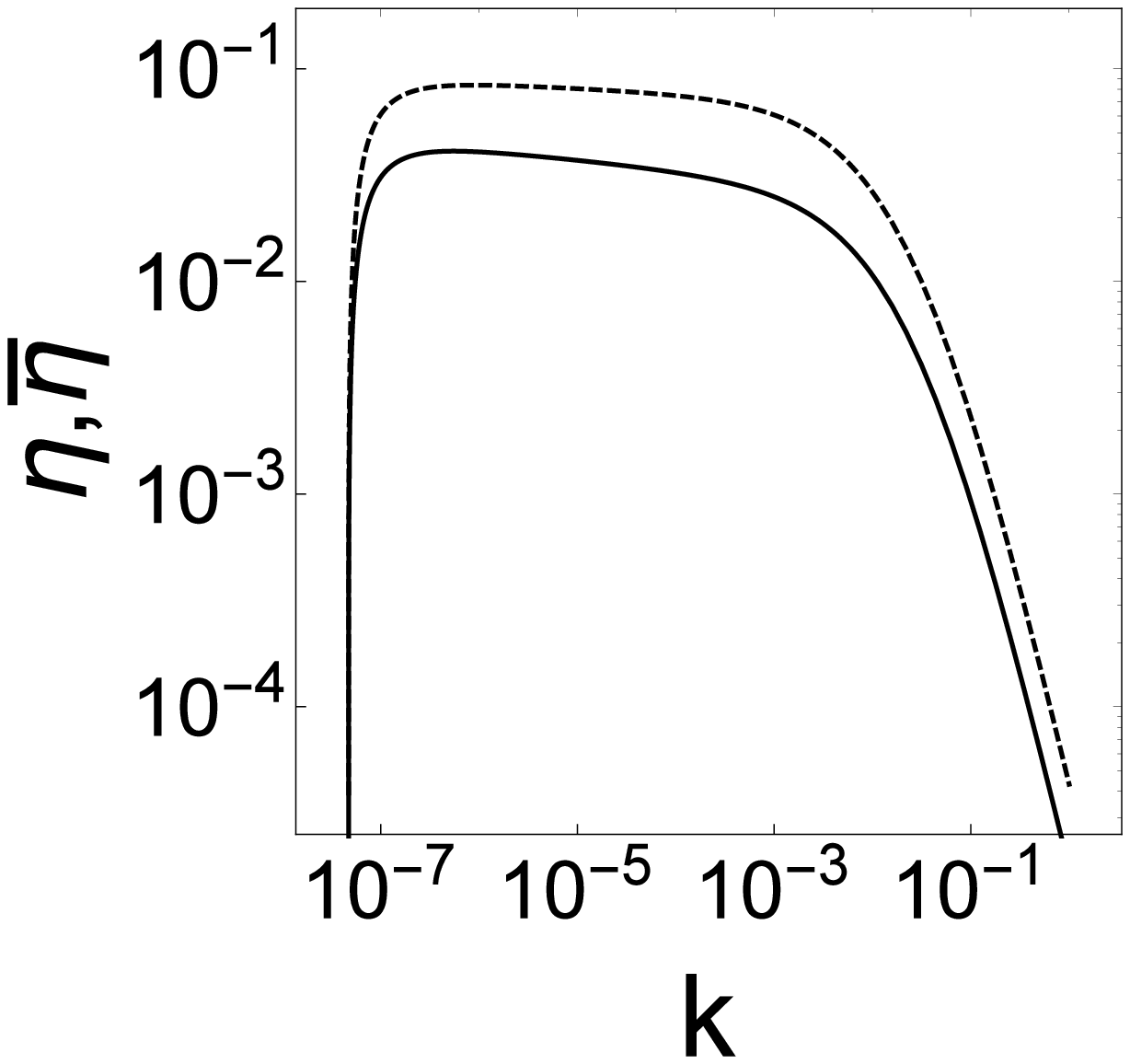,height=3.52cm,width=4.04cm,angle=0}
}
\caption{\label{fig:scdepNNLO2} Scale-dependences of the various couplings for the 3-dimensional $O(2)$ model  on the nearly critical trajectories in the NNLO6 approximation, evaluated by  Procedure B. The plot at the bottom to the right shows the scaling of the anomalous dimensions $\b{\eta}$ of the radial mode
(dashed line) and $\eta$ of the Goldstone modes (solid line).
}
\end{figure}

Since our trial to enforce $O(N)$ symmetry of the critical theory by setting $\h{z}=1$ has failed, we repeated the analysis of the RG evolution with  Procedure B, when the ratio $\h{z}$ was also  evolved. The typical scale-dependences are illustrated for the case $N=2$  in Fig.   \ref{fig:scdepNNLO2}. Even now a clearcut identification of the crossover scaling region can be made independently of the value of $N$ in which all investigated quantities  exhibit   power-law scale-dependences $ak^\alpha$ instead of keeping constant values. It is remarkable that the ratio $\h{z}$ increases strictly monotonically with decreasing scale $k$ and evolves to the peak value $\h{z}(k_c)$  significantly larger than 1 when the lower end of the WF crossover region is reached at the scale $k_c$ where $\b{\kappa}(k_c)=0$ sets on. At the same scale all the other investigated couplings as well as the anomalous dimensions take their maximum values. It should also be mentioned that for $N>3$ negative values of the scale-dependent coupling
$\b{Y}_\pe$ occur even now but in a rather narrow (one-order-of-magnitude wide) region before the power-law like,  positive valued WF scaling region is reached with decreasing scale $k$, the WF crossover region extends through 3 to 4 orders of magnitude in $k$. This seems to be an advantage of  Procedure B as compared to Procedure A.

The numerical values of the parameters $a$ and $\alpha$ characterizing the
power-law scalings of the various quantities were determined by fitting the scale-dependences in the WF crossover region and are given
 in Table  \ref{tab:on_zc2}. Since the modulus of all exponents $\alpha$ are much smaller than 1 and $k_c$ is less than $10^{-7}$, the values of the parameters $a$ can be identified with the fixed-point values. This is the case indeed. 
The educated guess for the fixed-point  values of the couplings needed to start the Newton-Rhapson algorithm for the search of the FP has been read off at the lower end of  the power-law scaling region at the scale $k_c$.
(For the   coupling $\b{\kappa}$ the power-law region ends up at the scale $k_l$ somewhat larger but of the same order of magnitude as $k_c$.)   It has been established that there exist  FPs for these values $\h{z}^*=\h{z}(k_c)$ for any $N$. The fixed-point values reproduced the values of the corresponding parameters $a$ with high precision. 
It has also been  found that the ratio $\h{z}$ scales as $\h{z}(k) = \h{z}^* k^{-[\b{\eta}(k_c)-\eta(k_c)]}$ in accordance with   Eq. \eq{bhzN} in the WF crossover region.

\begin{table*}[htb]
\begin{center}
\begin{tabular}{|c|c|c|c|c|c|c|c|c|}
\hline
 $N$   &  $\hat{z}^*$  & $\kappa^*$ & ${\b{\lambda}}^*$   & $10^4\b{Y}_\pa$      & $10^4\b{Y}_\pe$  & $\b{\eta} $ & $\eta$         \cr 
\hline\hline
$2$   & $1.65$ & $0.05k^{0.025}$   & $4.1 k^{-0.056}$    &$1.3 k^{-0.12}$  &$1.6 k^{-0.074}$    & $0.094 k^{-0.06}$    &  $0.020k^{-0.048}$             \cr
$3$   & $2.27$ & $0.06 k^{0.013}$   & $4.3 k^{-0.038}$    &$1.4 k^{-0.15}$  &$0.6 k^{-0.11}$    & $0.073 k^{-0.014}$    &  $0.026 k^{-0.021}$             \cr
$5$   & $2.78$ & $0.09 k^{0.008}$   & $3.7 k^{-0.024}$    &$6.0 k^{-0.08}$  &$ 0.1 k^{-0.19}$    & $0.084k^{-0.008}$    &  $0.028 k^{-0.008}$             \cr
$10$   & $3.5$ & $0.17 k^{0.004}$   & $2.4 k^{-0.011}$    &$160 k^{-0.037}$  &$0.08 k^{-0.2}$    & $0.093k^{-0.003}$    &  $0.022k^{-0.037}$             \cr
$100$   & $5.0$ & $1.9 k^{0.011}$   & $0.3 k^{-0.001}$    &$370k^{0.004}$  &$0.1 k^{-0.05}$    & $0.11 k^{0.01}$    &  $0.0032 k^{0.08}$             \cr
\hline
\end{tabular}
\end{center}
\caption{\label{tab:on_zc2} 
Scaling laws of the form  $a k^\alpha$ for the various quantities in the WF crossover region for the 3-dimensional $O(N)$ models found in the NNLO6 approximation in  Procedure B. The coefficients $a$
agree with high accuracy with the fixed-point values.
}
\end{table*}

Making use  of the power-law dependences of the various investigated quantities in the crossover scaling region,
one may start  the Newton-Rhapson algorithm with initial values
 $\h{z}(k_s)$, $\b{\kappa}(k_s)$, etc. taken at some intermediate scale
 $k_s\in [k_c,k_u]$,  where $k_u$ is the scale at which the WF scaling region starts, i.e., the smallest value
 at which it holds yet $\h{z}(k_u)=1$.  We have found numerically, on the one hand, that in  any of these cases the Newton-Rhapson algorithm reproduces with high accuracy the initial values as fixed-point values. For $k_s=k_u$ these fixed-point values are those obtained in Procedure A. On the other hand, the inconsistency of the used RG scheme, i.e., the nonvanishing of the beta-function $|\beta_{\h{z}}(k_s)|\sim\ord{\h{z}(k_s)\b{\eta}(k_s)}$ is present for all scales  $k_s$
and for any $N\ge 2$, but it has been found  minimal for $k_s=k_u$, i.e., essentially for Procedure A.
Therefore we see that in the parameter space there is a quasi-fixed line the points of which can be parameterized by the  values $\h{z}^*=\h{z}(k_s)$. The term quasi-fixed  line refers to the presence of increasing inconsistency with the choice of decreasing  $k_s$ values. Thus, insisting upon the principle of minimal inconsistency, i.e., that of minimal explicit breaking of the $O(N)$ symmetry of the critical theory, one has to identify the best estimates for the parameters of the WF FP as those obtained in Procedure A.

\subsection{Asymptotics for large $N$}

\begin{table}[htb]
\begin{center}
\begin{tabular}{|c||c|c||c|c|}
\hline
\multicolumn{1}{|c||}{}& \multicolumn{2}{|c||}{$\h{z}_\infty=1$} &  \multicolumn{2}{|c|}{$\h{z}_\infty=5$} \cr
\cline{2-5} 
    & Eq. \eq{asform} &  NNLO6u & Eq. \eq{asform}  &  NNLO6u \cr
\hline
$\kappa_\infty$  & $0.017$ & $0.0168$  & $0.017$  & $0.019$  \cr
$\lambda_\infty$  & $29.6$ & $29.0$  & $29.6$   & $30.0$  \cr
$\b{\eta}_\infty$  & $0.10$ & $0.099$  & $0.020$  & $0.11$  \cr
$\eta_\infty$  & $0.20$ & $0.20$  & $0.074$  & $0.32$  \cr
\hline
\end{tabular}
\end{center}
\caption{\label{tab:ascomp} Comparison of the asymptotic behaviour of our NNLO6u results with those
evaluated on the asymptotic formulas in Eq. \eq{asform} for the procedures
$\h{z}=1$ and $\h{z}=\h{z}^*$. The constants given in the table were obtained
from our NNLO6u values for $N=100$.
 }
\end{table}

The investigation of the behaviour of the WF FP for asymptotically large $N$ also hints to that the critical theory obtained in Procedure A should be favoured as the physically realistic one. On the base of the numerical NLO results obtained by Procedure A one may suggest that the asymptotic behaviour for large $N$ is given by the rules
\bea\label{largeN}
 && \b{\kappa}^*\sim\kappa_\infty N,~~\b{\lambda}^*\sim\frac{\lambda_\infty}{N},~~
 \eta^*\sim\frac{\eta_\infty}{N},
~~ \b{\eta}^*\sim\b{\eta}_\infty ,~~\h{z}^*=z_\infty,\nn
\eea
where the star refers to the values taken at the scale $k_c$ at which the WF crossover region ends. Inserting the assumptions \eq{largeN} into the  fixed-point equations $\beta_{\b{\kappa}}=\beta_{\b{\lambda}}=0$ and into the expressions \eq{andb} and  \eq{and}  for the anomalous dimensions, one finds that they are satisfied in the leading order of $N$ with the constants
\bea\label{asform}
&&\kappa_\infty=\frac{2\alpha_d}{d(d-2)},
~~ \lambda_\infty= \frac{d(4-d)}{4\alpha_d},\nn
&&\b{\eta}_\infty = \frac{ (4-d)^2 }{2 (d+2)z_\infty},
~~\eta_\infty=\frac{ 2(4-d)^2}{d^2-4} z_\infty g_\infty^2( d-2z_\infty g_\infty),\nn
&&g_\infty=\biggl( z_\infty +\frac{ 4-d}{d-2} \biggr)^{-1}
\eea 
for arbitrary value of $\h{z}^*=\h{z}(k_c)=z_\infty$.

We checked these asymptotic relations on the base of our numerical NNLO results obtained for $N=100$ and $d=3$ in Procedure A with $z_\infty=1$ and in Procedure B with $z_\infty=5$. The corresponding values of the various constants occurring in the asymptotic relations \eq{largeN} are given in Table  \ref{tab:ascomp}.
The comparison shows that Procedure A yields  NNLO6 values for the anomalous dimensions in agreement with the analytically predicted large $N$ asymptotic behaviour,  while there occurs some discrepancy for Procedure B.

\section{Summary}\label{sum}

The anomalous dimension and the correlation length's critical exponent characterizing the Wilson-Fisher (WF) fixed point (FP) in $O(N)$ models have been determined by means of the effective average action (EAA) renormalization group approach with the inclusion of the quartic gradient term of $\ord{\partial^4}$, using Litim's optimized regulator and field-independent derivative couplings.
For the $O(N)$ models with $N\ge 2$ our ansatz for the EAA enables one to evolve the derivative couplings of the radial and the Goldstone modes of the field independently.  
The next-to-next-to-leading-order (NNLO) effect has been evaluated as the
relative change  of the critical exponents comparing their NNLO and next-to-leading-order (NLO) values. The used RG framework is tested on the $O(1)$ model for dimensions $2<d<4$. We have established that the UV irrelevant higher-derivative coupling $\b{Y}$ becomes relevant at the WF FP. On the example of the 3-dimensional $O(1)$ model we studied the ${\dot R}$f effect, i.e., the neglection of the
scale-derivatives of the derivative couplings in the expression of the scale-derivative of the cutoff function. It was found that the NNLO effect and the  ${\dot R}$f effect are comparable when the modification of the NLO values of the critical exponents are considered, but the  ${\dot R}$f effect becomes with an order of magnitude weaker when the NNLO approximation is used. For the $O(1)$ model for dimensions $d=4-\epsilon$ it is shown that the higher-derivative coupling vanishes proportionally to $\epsilon^3$ in the limit $\epsilon\to 0$ of the $\epsilon$-expansion.
It is found that the validity range of the RG scheme used by us is restricted to the dimensions $d$ in the interval $2.7<d<4$, otherwise the EAA of the critical theory becomes unbounded from below due to the negative sign of the higher-derivative coupling $\b{Y}$. The NNLO effect has been determined in the validity range of the RG scheme.

 The $N$-dependence of the NNLO effect on the critical exponents of the WF FP of the 3-dimensional $O(N)$ models for $N\ge 2$ was determined. It was shown that the treatment of the ratio $\h{z}$ of the wavefunction renormalization of the radial mode to that of the Goldstone modes is rather crucial when the WF crossover scaling is considered. Keeping $\h{z}=1$, i.e., keeping the $O(N)$ symmetry
of the EAA at the NLO level yields crossover regions even in the NNLO scheme  in which the characteristic quantities of the WF FP are constant, while letting $\h{z}$ evolve leads to power-law like scalings. It was shown, on the one hand, that the fixed-point equations have solution for any given value of the ratio $\h{z}$. On the other hand, we have found that  the applied RG schemes show up the inconsistency that the
beta-function of the ratio $\h{z}$ does not vanish exactly, when calculated from the fixed-point solutions. It was shown, however, that this inconsistency minimizes for $\h{z}=1$  providing a critical theory which preserves $O(N)$ symmetry with the accuracy of the order of the anomalous dimension of the radial mode of the field.  Therefore the RG scheme with keeping $\h{z}=1$ constant is favoured as the physically realistic one. The  $N$-dependence of the anomalous dimension of the Goldstone modes and that of the correlation length's critical exponent have been found for $\h{z}=1$ qualitatively similar to those of known NLO results.
 The NNLO effects on these critical exponents have been determined as the function of $N$. It was also shown that the coupling of the quartic derivative term of the Goldstone modes takes negative critical values for $N>3$ and makes the EAA unbounded from below.

\section*{Acknowledgements}
S. Nagy acknowledges financial support from a J\'anos Bolyai Grant of the
Hungarian Academy of Sciences, the Hungarian National Research Fund OTKA
(K112233) and Z. Peli acknowledges the support through the new National Excellence Program  of the Ministry of Human Capacities.

\appendix

\section{Short overview on the present status of the determination of the critical exponents of the WF FP in the functional RG framework}\label{app:over}

It is worthwhile to outline a rather short overall picture of the present 
 status of the determination of the exponents $\nu$ and $\eta$ at the phase
 transition point of the  $O(N)$ models by means of various functional RG
 methods. The critical exponents $\nu$ and $\eta$ are universal and -as it is well-known -  the $O(N)$ models with $N=1$, $N=2$, and  $N=3$ belong to the Ising, XY-, and Heisenberg universality class, respectively. The determination of their universal critical exponents has been one of the intensive and successful applications   of the various functional RG methods  such as the RG frameworks proposed by Wilson \cite{Wilso1974}, Wegner and Houghton \cite{Wegne1973}, Polchinski \cite{Polch1984}, Wetterich \cite{WePLB1993},   Blaizot-Mendez-Wschebor(BMW) \cite{Blaiz2006a,Blaiz2006b}, and the proper-time RG method \cite{Olesz1994,Flore1995,Liao1996}. Most of the investigations of the universal critical exponents of the $O(N)$ models were performed in some truncated forms of the GE, an approximation scheme that relies on the smallness of the wavefunction renormalization \cite{Morri1994,Litim2001} and has been applied successfully in a great variety of physical problems (see e.g. \cite{Wette2002}). The GE is particularly applicable to the determination of the quantities defined at vanishing momenta such as the critical exponents and the phase diagrams, but it does not allow to find the full momentum-dependence of the correlation functions.
As opposed to it the BMW method \cite{Blaiz2006a,Blaiz2006b} relies on the flow of the one-particle irreducible (1PI) vertices and enables one to take with their  full  momentum-dependence properly \cite{Benit2009,Benit2011}. The GE has been applied  in the LPA, the NLO, and the NNLO of the GE. The  BMW approximation scheme at the order $s$ consists of setting the internal momenta to zero in all 1PI vertices of order larger than $s$ and achieving a closed set of flow equations for the first $s$ 1PI vertex functions.
 Beyond the determinations of the critical exponent $\nu$ in the LPA, great efforts have been made to determine $\nu$ and  $\eta$ in the NLO of the GE taking with the evolution of either the uniform (NLOu) \cite{Tetra1993} or the field-dependent (NLOf) \cite{Tetra1993,Morri1994,BallH1995,Berge1997,Comel1998,Seide1999,Mazza2001,Canet2002,Canet2003,Wette2002,Ballh2004,Zappa2012}  wavefunction renormalization. Recently results have been obtained in NNLO of the GE \cite{Canet2003} and in BMW with full momentum dependence up to the order $s=2$ \cite{Benit2011}. In \cite{Canet2003} the field-dependence of the gradient terms have been properly taken into account (NNLOf) after a thorough discussion of its truncation.

 It is well-known that the NLO results  (especially the NLO values of the anomalous dimension $\eta$)  depend on the renormalization scheme, the choice and the parameter(s) of the regulator, but the optimization by applying the principle of minimal sensitivity \cite{SenBe2000,Litim2001b} or requiring reparametrization invariance of physical quantities \cite{Wette2002} can remove the ambiguity caused by the choice  of the regulator parameter(s). In Tables \ref{tab:critexplit}
 and \ref{tab:critexplitN}
we listed the values of  the anomalous dimension $\eta$ and the correlation length's critical exponent $\nu$ for the $d=3$ dimensional $O(1)$ and $O(N\ge 2)$ models published in the recent years (without pretending to be complete), concentrating  on the NLO and NNLO results of the GE. We also list the best available BMW results \cite{Benit2009,Benit2011} and some results obtained by other methods \cite{Guida1998,Pogor2008,Hasen2010,Campo2002,Campo2006,Hasen2001JPA,Anton1995,Moshe2003} that can be considered  the `world's best' values, as well as experimental values for the Ising model \cite{Mulle1999}.
Before turning to a discussion of  Tables \ref{tab:critexplit}
and \ref{tab:critexplitN}, it is worthwhile to emphasize that the
results for the anomalous dimension are rather sensitive to the field-dependence of the wavefunction renormalization and that of the coupling of the $\ord{\partial^4}$ term (see a detailed discussion in \cite{Canet2003}).  The field-dependence of the wavefunction renormalization was found to show up rather specific features in the IR limit \cite{Morri1994,Seide1999,Conso2006,Zappa2012}.

The  results listed in  Tables \ref{tab:critexplit} and \ref{tab:critexplitN}  were obtained in the following RG frameworks:
\begin{itemize}
 \item in \cite{Gersd2001,Wette2002} Wetterich's EAA RG method with an exponential regulator was used enabling a formulation which respects reparametrization invariance of physical quantities; the field-dependence of the potential and  that of the wavefunction renormalization was approximated by polynomials of order third and first orders, respectively, of the variable $\rho=\hf \phi^2$ with expansion around the nontrivial minimum of the potential;
\item in \cite{Canet2002,Canet2003} the EAA RG method with PMS optimization  for one-parameter families of exponential and Litim type regulators \cite{Litim2001b} was used and good convergence achieved for field-dependent potential and derivative couplings of high order;
\item in \cite{Mazza2001} the proper time RG method \cite{Olesz1994} with   a  one-parameter family of proper time regulators  was used and an optimized limit found by sharpening the effective width of the regulator, the field-dependence of the FP and the scaling functions at the FP  were determined both in LPA and NLO;
\item in \cite{Seide1999}  Wetterich's EAA RG method with an exponential regulator was used, the full field-dependence of both the critical potential and  wavefunction renormalization were approximated by higher-order polynomials plus their asymptotic expressions;
\item in \cite{Comel1998} Polchinski's RG method with  a one-parameter family of generalized Lorentzian regulator functions was used and optimal values were determined after a rather involved discussion,  the full field-dependence of both the critical point and the scaling functions was obtained;
\item in \cite{Morri1994} flow equations for the Legendre effective action
(determining the  1PI part of the Wilson effective action \cite{Morri1993IJMPA}) was used
with a  power-law regulator and the full field-dependence of both the critical point and the scaling functions  obtained;
\item in \cite{Tetra1993} and \cite{Berge1997} the EAA RG method was used with an exponential regulator and uniform wavefunction renormalization;
\item in \cite{Benit2009,Benit2011} the BMW method with a one-parameter family of exponential regulators was used with PMS optimization.
\end{itemize}

 \begin{table}[htb]
\begin{center}
\begin{tabular}{|c|c|c|c|c|}
\hline
 Approximation  & Ref. &$\eta$ & $\nu$ \cr
\hline \hline 
  &  & & \cr
 NLOf    & \cite{Wette2002}    & 0.0467 & 0.6307 \cr
     & \cite{Canet2002}    & 0.0443 & 0.6281\cr
     &                     & 0.0470 &  0.6260\cr
     & \cite{Mazza2001}    & 0.0330  & 0.6244 \cr
     & \cite{Seide1999}    & 0.0467 & 0.6307\cr
     &\cite{Comel1998}     & 0.042  & 0.633 \cr
     & \cite{Morri1994}    & 0.05393  & 0.6181\cr
  & & & \cr
\hline
  &  & & \cr
NLOu     &\cite{Tetra1993}     & 0.045 & 0.638 \cr
  & & & \cr
\hline 
 &  &&\cr
 NNLOf & \cite{Canet2003}    & 0.033 &  0.632 \cr 
\hline
& & & \cr
 BMW & \cite{Benit2009,Benit2011}& 0.039 & 0.632\cr
\hline
&&& \cr
MC    &\cite{Hasen2010} &  0.03627(10) & 0.63002(10)\cr
FT    & \cite{Pogor2008} & 0.0318(3)     &  0.6306(5)  \cr
7-loop & \cite{Guida1998} & 0.0335(25) &   0.6304(13) \cr
Exp.  & \cite{Mulle1999}  &0.045(11)&0.636(31)  \cr
& & & \cr
\hline  
\end{tabular}
\end{center}
\caption{\label{tab:critexplit} The critical exponents for the $O(1)$ model for $d=3$ obtained by various RG methods. For comparison the `world's best' values of the critical exponents of the Ising-model for $d=3$ are also listed from Monte Carlo simulations \cite{Hasen2010}, field theoric RG with summation of divergent series \cite{Pogor2008}, from perturbation theory including 7-loop contributions \cite{Guida1998}, and from experiment in mixing transition \cite{Mulle1999}.
  }
\end{table}
\begin{table}[htb]
\begin{center}
\begin{tabular}{|c|c|c|c|c|}
\hline
$N$& Approximation  & Ref. &$\eta$ & $\nu$ \cr
\hline \hline 
 &   & & &\cr
 2 &NLOf &\cite{Gersd2001,Wette2002}     & 0.049 & 0.666 \cr
   &NLOu& \cite{Tetra1993}   &0.042 & 0.700\cr
   &BMW &  \cite{Benit2011}    & 0.041 & 0.674 \cr 
   \cline{2-5}
   &MC& \cite{Campo2006}      &0.0381(2) &0.6717(1)\cr
   &FT & \cite{Pogor2008}     & 0.0334(2)&0.6700(6)\cr
 &&&&\cr
\hline
&&&&\cr
 3 &NLOf &\cite{Gersd2001,Wette2002}     & 0.049 &0.704\cr
   & NLOu & \cite{Tetra1993}   &0.038 & 0.752 \cr
   &BMW & \cite{Benit2011}   & 0.040   & 0.715\cr
 \cline{2-5}
  &MC&  \cite{Campo2002}     &0.0375(5) &0.7112(5)\cr
   &FT & \cite{Pogor2008}     & 0.0333(3)&0.7060(7)\cr
 &&&&\cr
\hline
&&&&\cr
 4     &NLOf &\cite{Gersd2001,Wette2002}     & 0.047 &0.739\cr
    & NLOu &  \cite{Tetra1993}   & 0.034 & 0.791\cr
    &BMW & \cite{Benit2011}  & 0.038    & 0.754\cr
  \cline{2-5}
  &MC& \cite{Hasen2001JPA} & 0.0365(10)& 0.749(2)\cr
    &7-loop &  \cite{Guida1998}    &0.0350(45)&0.741(6)\cr
 &&&&\cr
\hline
&&&&\cr
 10       &NLOf &\cite{Gersd2001}     & 0.028 &0.881\cr
      &NLOu &  \cite{Tetra1993}   & 0.019 & 0.906\cr
      &BMW & \cite{Benit2011}  & 0.022    & 0.889\cr
 \cline{2-5}
      &6-loop & \cite{Anton1995} & 0.024 &0.859\cr
&&&&\cr
\hline
&&&&\cr
100  &NLOf &\cite{Gersd2001}     &0.0030 & 0.990\cr
     &NLOu &  \cite{Tetra1993}   &0.002 & 0.992\cr
    & BMW &\cite{Benit2011}  &0.0023  & 0.990   \cr
 \cline{2-5}
     &$1/N$-exp. &\cite{Moshe2003}  & 0.0027 & 0.989 \cr
&&&&\cr
\hline  
\end{tabular}
\end{center}
\caption{\label{tab:critexplitN} The critical exponents for the $O(N)$ model for $d=3$ obtained by various RG methods.  For comparison a few recent results are also listed from Monte Carlo simulations\cite{Campo2006,Campo2002,Hasen2001JPA}, field theoric RG with summation of divergent series \cite{Pogor2008}, from perturbation theory including higher-loop contributions \cite{Guida1998,Anton1995},and from  large $N$ limit of quantum field theory in the $1/N$ expansion  \cite{Moshe2003}.}
\end{table}

It can be observed in Table \ref{tab:critexplit} that the values of  the critical exponent $\nu$ for $N=1$ and $d=3$ obtained in the NNLOf and NLOf of the GE in the frameworks of various RG schemes agree within a few per cents, while  those of the anomalous dimension $\eta$ may differ by several 10 per cents. In a given RG scheme, however, generally the values of $\eta$ increase and those of $\nu$ decrease when the field-dependence is properly taken into account: for example,
the increment of $\eta$ is $\sim 4$ per cents and the decrement of $\nu$ is
$\sim 1$ per cents comparing the NLOf results of \cite{Wette2002} to the
 NLOu results of \cite{Tetra1993}. The importance of the effect of taking into account the evolution of the term of $\ord{\partial^4}$ of the EAA reveals itself in comparison of the NLOf data of \cite{Wette2002} and the NNLOf data of 
\cite{Canet2003}; it makes out a decrement of $\sim 30$ per cents  and
an increment of $0.2$ per cents of the NLOf values of $\eta$ and $\nu$, respectively.  It is worthwhile mentioning that the NNLOf data of \cite{Canet2003} are in excellent agreement with the BMW values \cite{Benit2009,Benit2011} and with
the  the world's best estimates.

In Table \ref{tab:critexplitN} we listed the most recent values of $\eta$ and $\nu$ for various $O(N)$ models with $N\ge 2$ for the number of dimensions $d=3$. One can see, on the one hand,  that the NLOf values for $\eta$ are generally larger than the NLOu values and the latter are closer to the BMW values and the world's best other estimates.  Nevertheless, the NLOu data show similar qualitative behaviour in their $N$-dependencies as the NNLO data.
Moreover the anomalous dimension decreases to zero whereas the critical exponent saturates at $\nu=1$ for $N$ increasing to infinity, in agreement with field theoretic expectations \cite{Moshe2003}.  According to our knowledge there are no available NNLO data neither in the NNLOu nor in the NNLOf approximations which
would  enable one to make conclusions on the significance and the $N$-dependence of the NNLO effect on the values of the anomalous dimension $\eta$ and the critical exponent $\nu$.

For the $O(N)$ models for dimensions $d=4-\epsilon$ the leading order results from the $\epsilon$ expansion are given as
\bea\label{2loop}
  \b{\lambda}^*_K &=&\frac{1}{\alpha_4}\biggl(\frac{3}{N+8}\epsilon +\ord{\epsilon^2}\biggr),\nn
 \eta&=& \frac{N+2}{2(N+8)^2 } \epsilon^2+\ord{\epsilon^3},\nn
 \nu&=& \frac{1}{2} +\frac{N+2}{4(N+8)}\epsilon +\ord{\epsilon^2},
\eea
where $\b{\lambda}^*_K$ is the critical  coupling of the quartic term $\phi^4$ \cite{Klein2001} (related to our definition of coupling $\b{\lambda}^*_K=3\b{\lambda}^*$). One can see that the  position of the WF FP and the critical exponent $\nu$ depend linearly, the anomalous dimension $\eta$ depend quadratically on  $\epsilon$ in the limit $\epsilon\to 0$  (see e.g. \cite{Wilso1974,Guida1998,Klein2001}).

\section{RG evolution equations in the approximation NNLO$2\eta$ for the $O(1)$ model}\label{app:nnlo2e}

Taking the ${\dot R}$f effect into account means that one has to insert the full expression \eq{rdoteta} of  the scale-derivative of the cutoff function into the flow Eqs. \eq{potflow} and \eq{wfrflow}. Then in the approximation NNLO$2\eta$ one  arrives after lengthy manipulations to the following evolution equations for the dimensionless couplings,
\begin{widetext}
\bea\label{bkappannloe}
\dot{\b{\kappa}}
&=&
-(d-2+\eta)\b{\kappa}+a\biggl(1+2\b{Y}_k-\frac{\eta}{d+2}
+2\frac{\beta_{\b{Y}}-(2+\eta)\b{Y}}{d+4}\biggr)\b{g}^2\equiv\beta_{\b{\kappa}},
\eea
\bea\label{blambnnloe}
\dot{\b{\lambda}}
&=&
(d-4+2\eta)\b{\lambda}+b\biggl(1+2\b{Y}_k-\frac{\eta}{d+2}
+2\frac{\beta_{\b{Y}}-(2+\eta)\b{Y}}{d+4}\biggr)
\b{\lambda}^2\b{g}^3 \equiv\beta_{\b{\lambda}},
\eea
\bea\label{bYnnloe}
{\dot{\b{Y}}}
&=& 
\frac{2 A_2- A_2 B_1 +A_1 B_2+(32 - 8A_1  -16  B_1 -A_2 C_1  - 2 B_2 C_1 -4 C_2  +A_1 C_2  +2 B_1 C_1 )\b{Y}}{16- 8B_1 - B_2 C_1 - 2C_2 + B_1 C_2}
\equiv\beta_{\b{Y}} ,
\eea
\end{widetext}
and the expression for the anomalous dimension
\bea\label{andnnloe} 
\eta 
&=& 
-\frac{8 A_1+A_2 C_1- A_1 C_2}{16- 8B_1 - B_2 C_1 - 2C_2 + B_1 C_2}.
\eea
The coefficients $A_i,~B_i,~C_i$ $(i=1,2)$  are given as
\bea
&&A_1 = (2+ 4 \b{Y}) I(0),~~
A_2 = (2+ 4 \b{Y}) J(0), \nn
&&B_1 = I(0)-I(1),~~
B_2 = J(0)-J(1),\nn
&&C_1 = I(0)-I(2),~~
C_2 = J(0)-J(2)
\eea  
through the (dimensionless) loop-integrals
\bea
I(2n)&=&k^{-2n} \int_p p^{2(n-1)}\partial_{Q}^2\Phi (Q;r,p^2)\biggr|_{Q=0},
\nn
J(2n)&=&k^{-2n} \int_p p^{2(n-2)}\partial_{Q}^4\Phi (Q;r,p^2)\biggr|_{Q=0}
\eea
in terms of the  function
\bea
\!\!\!\! \Phi (Q;r,p^2) &=& 2r^*[ 2r^* U_k'''(r^*) + 3 U''(r^*)]^2 \nn
&&\times\biggl( G^2(p^2) [G(q^2)]_{q=Q-p }-  G^3(p^2)  \biggr).
\eea
The notations $\partial_{Q}^2$ and $\partial_{Q}^4$ are symbolic, second and fourth partial derivatives with respect to the Euclidean momentum $Q_\mu$  are meant followed by the replacements given in Eq. \eq{odreps}. The explicit forms of the loop integrals are given as 
\begin{widetext}
\bea 
I(2n)
&=& 
18 \alpha_d \b{\kappa} \b{\lambda}^2 \bar{g}^4\biggl[ \biggl(\frac{4}{d (d+2+2n)}+\frac{32 \bar{Y}}{d (d+4+2n)}+\frac{96 \bar{Y}^2}{d (d+6+2n)}\biggr) \b{g} -\frac{6  \bar{Y}}{d+2+2n} -\frac{12 \bar{Y}}{d (d+2+2n)}-\frac{1}{d+2n}\biggr],
\nn
J(2n)
&=&
72 \alpha_d \b{\kappa} \b{\lambda}^2 \bar{g}^4
\biggl[ \biggr(\frac{1792 \bar{Y}^4}{d^2 (d+12+2n)}+\frac{6144 \bar{Y}^4}{d (d+2) (d+12+2n)}+\frac{1280  \bar{Y}^3}{d^2 (d+10+2n)}  +\frac{3840  \bar{Y}^3}{d (d+2) (d+10+2n)}
\nn
&&
+\frac{192 \bar{Y}^2}{d^2 (d+8+2n)}
+\frac{2304 \bar{Y}^2}{d  (d+2) (d+8+2n)} +\frac{576 \bar{Y}}{d (d+2) (d+6+2n)}+\frac{48\bar{Y}}{d (d+2+2n)}
+\frac{48}{d (d+2) (d+4+2n)}\biggr)\b{g}^3
\nn
&&
+\biggl(\frac{1152 \bar{Y}^3}{d^2 (d+8+2n)}+\frac{960  \bar{Y}^3}{d (d+8+2n)}
+\frac{2304  \bar{Y}^3}{d (d+2) (d+8+2n)}+\frac{640 \bar{Y}^2}{d^2 (d+6+2n)}
 +\frac{480 \bar{Y}^2}{d (d+6+2n)}
\nn
&&
+\frac{80 \bar{Y}}{d^2 (d+4+2n)}+\frac{120  \bar{Y}}{d (d+4+2n)}+\frac{12 }{d (d+2+2n)}\biggr)\b{g}^2
\nn
&&
+\biggl(\frac{48 \bar{Y}^2}{d^2 (d+4+2n)}+\frac{288  \bar{Y}^2}{d (d+4+2n)}+\frac{144  \bar{Y}^2}{d (d+2) (d+4+2n)}
+\frac{24 \bar{Y}^2}{d+4+2n}+\frac{8  \bar{Y}}{d+2+2n}+\frac{1}{d+2n}\biggr)\b{g}
-\frac{3 \bar{Y}}{d+2n}\biggr]
\nn
\eea
\end{widetext}
with $\b{g}$ given in Eq. \eq{bgNNLO}.

\end{document}